\documentclass[preprint,12pt]{elsarticle}

\usepackage{amssymb}
\usepackage{amsmath}
\usepackage{algorithm}      
\usepackage{algorithmic} 
\usepackage{multirow} 
\usepackage{graphicx}
\usepackage{subcaption}
\usepackage{url}
\usepackage{hyperref}
\usepackage{cleveref}
\usepackage{booktabs} 
\usepackage{float}
\usepackage{threeparttable}
\usepackage{xcolor}
\usepackage{enumitem}
\usepackage{booktabs}

\journal{Information Sciences}

\begin{document}

\begin{frontmatter}


\title{Adaptive Preference Modeling via Explicit Indirect Relational Learning for Personalized Fashion Matching}

\author{Shuiying Liao} 
\author{Li Li}
\author{P.~Y.~Mok\corref{cor1}} 
\cortext[cor1]{Corresponding author: tracy.mok@ust.hk}
\affiliation{organization={The Hong Kong University of Science and Technology},
            addressline={Clear Water Bay},
            country={Hong Kong}}

\begin{abstract}
Personalized fashion complementary recommendation requires jointly modeling user preferences and item compatibility under sparse and multimodal data conditions. Existing approaches often capture higher-order relational signals implicitly through graph propagation or rely on direct interaction data, limiting their ability to explicitly model indirect preference and compatibility relationships.
To address this limitation, we propose an \textbf{A}daptive \textbf{P}reference with \textbf{C}ontrastive \textbf{L}earning (APCL) framework that explicitly models both direct and indirect relational signals within a unified recommendation architecture. Specifically, APCL constructs indirect user–item and item–item relationships through a correlation-guided adaptive aggregation mechanism and represents them as dedicated personalization and compatibility views. To enhance representation learning, we further introduce a functional-view contrastive learning strategy that aligns direct and indirect preference representations as well as direct and indirect compatibility representations, encouraging consistency across relational contexts.
By integrating multimodal visual and textual information with explicit indirect relational modeling, APCL is able to capture richer semantic characteristics while improving robustness under sparse interaction settings. Experiments on two benchmark fashion recommendation datasets demonstrate that APCL consistently outperforms representative baseline methods, particularly in cold-start scenarios. The results suggest that explicitly modeling indirect relational structures and aligning them through contrastive objectives provides an effective approach for personalized complementary recommendation. 
\end{abstract}

\begin{graphicalabstract}
\includegraphics[width=\linewidth]{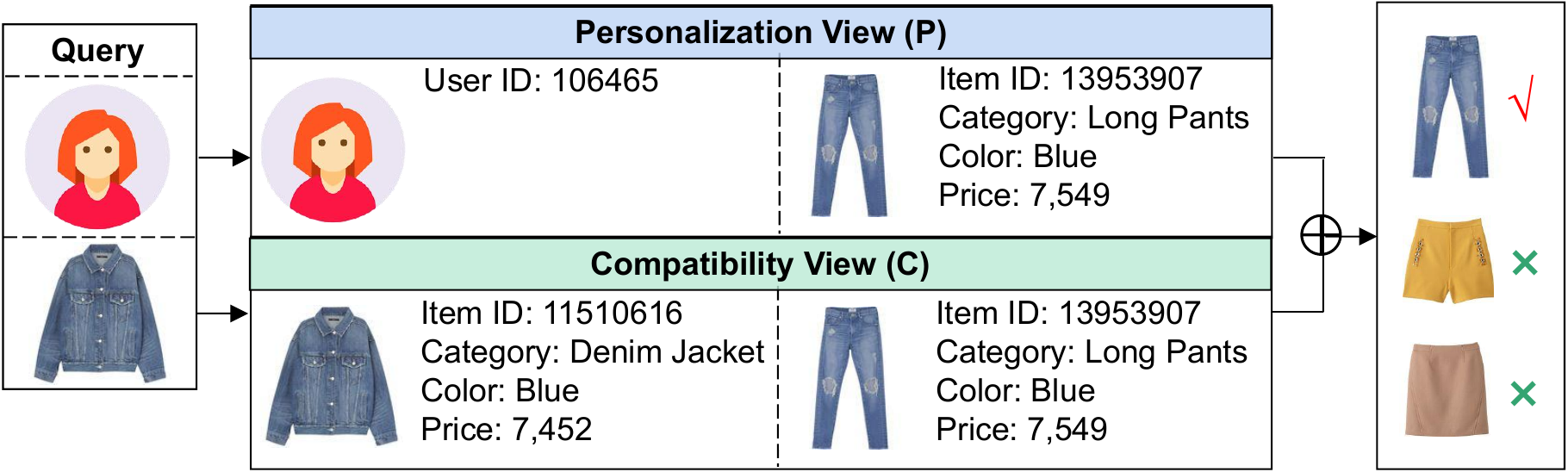}
\includegraphics[width=\linewidth]{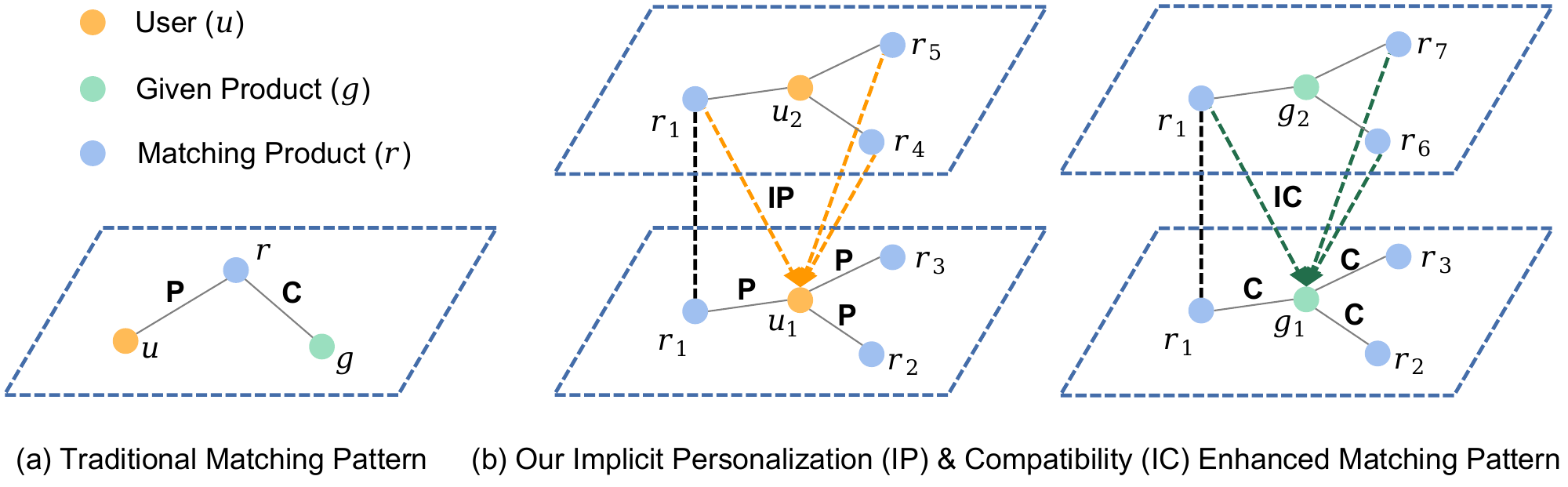}
\end{graphicalabstract}

\begin{highlights} 
\item Adaptive preference modeling framework based on explicit indirect relational learning. 
\item Explicit construction of indirect user--item and item--item relational views. 
\item Functional-view contrastive alignment of direct and indirect relational representations. 
\item Correlation-guided aggregation enhances learning from sparse interaction data. 
\item Effective and robust personalized fashion matching under cold-start conditions. 
\end{highlights}


\end{frontmatter}


\section{Introduction}
\begin{figure}[t]
  \centering
  \setlength{\abovecaptionskip}{0.2cm}
  \includegraphics[width=\linewidth]{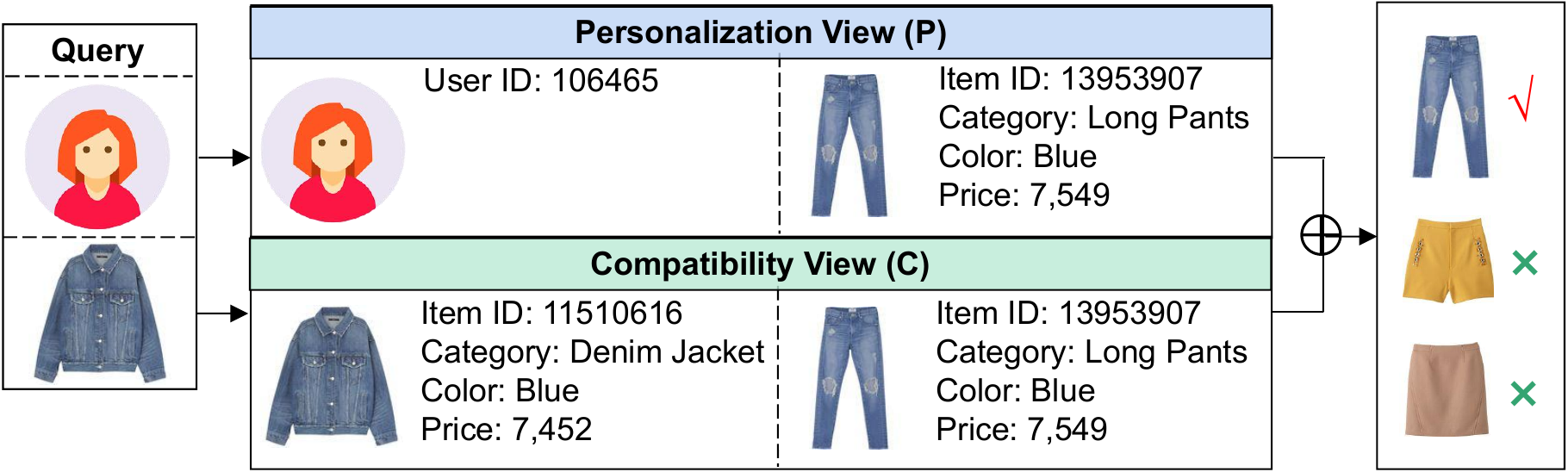}
  \caption{Personalized Fashion Complementary Recommendation Sample. Each transaction includes images, textual information, user IDs and product IDs.}
  \label{fig: trans}
\end{figure}

Personalized recommendation systems have become an indispensable component of modern e-commerce platforms, helping users discover products that align with their interests while improving user engagement and conversion rates~\cite{deldjoo2023review,raza2026comprehensive}. In the fashion domain, recommendation tasks exhibit unique characteristics compared with conventional item recommendation. Rather than predicting a user's preference for an individual product, fashion recommendation often involves identifying complementary items that can be combined into a coherent outfit, such as matching a bottom garment to a given top~\cite{GPBPR, CP}. Consequently, personalized fashion matching, as illustrated in \Cref{fig: trans}, requires the simultaneous modeling of two interdependent objectives: capturing the unique preferences of individual users (\textbf{personalization}) and identifying harmonious relationships among fashion products (\textbf{compatibility})~\cite{ma2024personalized, GPBPR, liu2024unifying}. Achieving both objectives simultaneously remains a challenging problem in recommendation research.

Existing approaches to personalized fashion matching can be broadly categorized into collaborative filtering, content-based, and hybrid recommendation methods. Collaborative filtering methods leverage user--item interaction histories to infer latent user preferences~\cite{BPR, he2020lightgcn}, while content-based approaches exploit product attributes, visual characteristics, and textual descriptions to model item similarity and compatibility~\cite{kang2019complete, de2015content}. More recently, multimodal recommendation methods have integrated heterogeneous information sources to capture fine-grained semantic characteristics of fashion products~\cite{ma2024personalized, huang2024multimodal}. In parallel, graph-based recommendation methods have demonstrated strong performance by modeling higher-order relationships through propagation over user--item interaction graphs~\cite{wang2019neural, he2020lightgcn}. These advances reflect a growing trend toward jointly modeling personalization and compatibility within unified representation learning frameworks~\cite{PCE, li2020hierarchical}.

Despite these developments, personalized fashion matching remains challenging for three primary reasons. \textit{First}, fashion preference is inherently personal and serves as an important medium of individual expression. Although compatibility-oriented recommendation methods can effectively model relationships among products, they often struggle to capture the subtle preference patterns embedded within user--item interactions~\cite{GPBPR, CP}. As a result, recommendations that appear compatible from a product perspective may not necessarily align with the preferences of individual users. 
\textit{Second}, fashion recommendation suffers from severe data sparsity. Due to the short lifecycle of fashion products and the rapid evolution of fashion trends, many items accumulate only limited interaction records before leaving the market. In the era of fast and ultra-fast fashion, retailers continuously introduce large volumes of new products, resulting in sparse user--item and item--item interaction data~\cite{gallery2024fashion}. This sparsity affects both personalization modeling and compatibility modeling, making it difficult to learn robust representations from direct observations alone~\cite{jing2023contrastive, wei2021contrastive}. \textit{Third}, fashion products are inherently multimodal. On e-commerce platforms, products are typically represented through diverse information sources, including images, textual descriptions, attributes, and category information~\cite{kang2019complete, deldjoo2023review}. Although these modalities provide rich semantic cues for recommendation, effectively integrating heterogeneous information remains a challenging problem, particularly when interaction data are sparse and supervision signals are limited~\cite{liao2024hypergraph, huang2024multimodal, liao2026consistency}.

\begin{figure}[t]
  \centering
  \setlength{\abovecaptionskip}{0.2cm}
  \includegraphics[width=.8\linewidth]{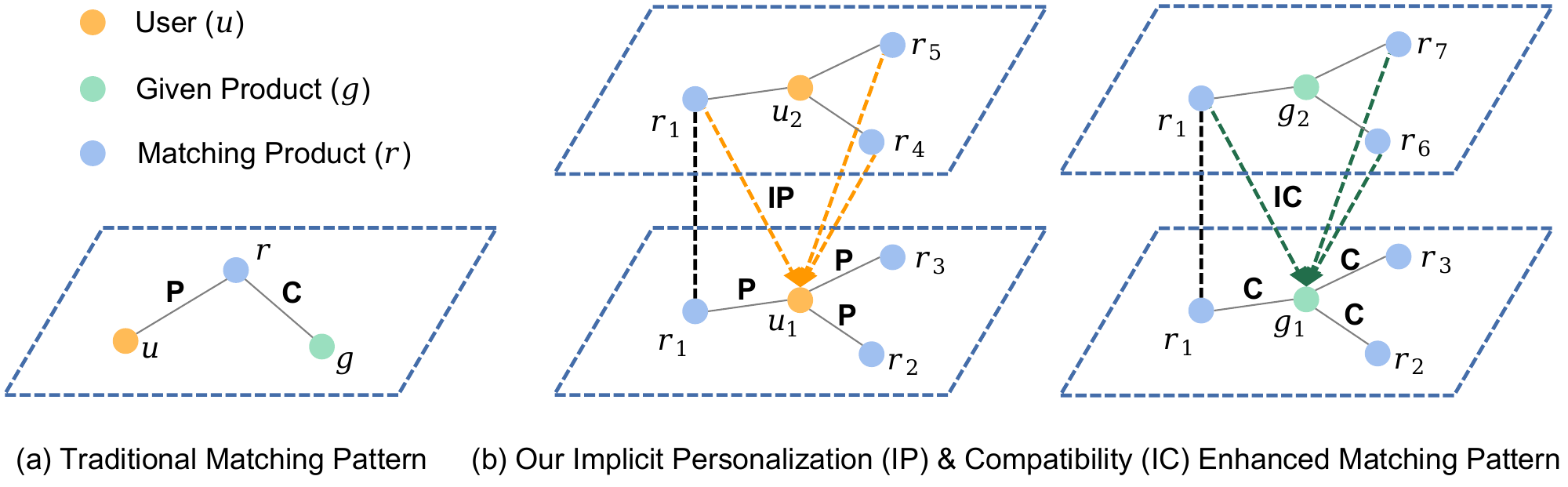}
  \caption{Personalized fashion recommendation modeling patterns comparison.}%
  \label{fig:example}
\end{figure}

A key observation motivating this work is that many informative recommendation signals are indirect rather than directly observed. For example, users who share similar preferences toward certain products may provide valuable information for understanding the latent interests of a target user, even when direct interactions are limited. Similarly, products that frequently appear within similar interaction contexts may reveal latent compatibility relationships beyond those explicitly observed in historical data. Such indirect relational signals are particularly valuable in sparse environments, where direct evidence is often insufficient to support reliable representation learning~\cite{wang2019neural, he2020lightgcn, liao2026hamiltonian, wei2021contrastive, cfalr}.

Existing recommendation approaches often capture these indirect relationships \textit{implicitly}~\cite{wang2024mmpgcn}. Graph-based collaborative filtering methods model higher-order relationships through iterative information propagation on interaction graphs~\cite{wang2019neural, he2020lightgcn}, while co-occurrence-based approaches infer latent associations from shared interaction statistics~\cite{ebesu2018collaborative, weijun2014collaborative}. Although effective, these approaches typically embed different types of relational signals within a unified representation space, making it difficult to explicitly distinguish, interpret, and control the contributions of indirect preference and compatibility relationships. Consequently, indirect relational signals are often treated as by-products of the learning process rather than as explicit modeling objectives.

To address this limitation, we propose \textbf{Adaptive Preference with Contrastive Learning (APCL)}, a framework for personalized fashion matching based on \textbf{explicit indirect relational learning}. Rather than relying solely on implicit propagation mechanisms, APCL explicitly constructs indirect user--item and item--item relationships through a correlation-guided adaptive aggregation strategy. These relationships are organized into two functional dimensions of recommendation, namely \textbf{personalization} and \textbf{compatibility}, and further represented through corresponding direct and indirect relational views. This formulation enables more structured modeling of relational information while preserving the complementary roles of preference learning and compatibility learning.

Building upon this formulation, we introduce a functional-view contrastive learning mechanism that aligns direct and indirect representations within both personalization and compatibility spaces. Unlike conventional contrastive recommendation methods that primarily focus on modality alignment, graph augmentation, or cross-view consistency~\cite{wu2022multi, zou2022multilevel, ma2022crosscbr}, the proposed approach encourages consistency across relational contexts. By treating indirect relational signals as an additional source of supervision, the model (see~\Cref{fig:example}) is able to improve representation robustness and generalization under sparse interaction settings~\cite{jing2023contrastive, wei2021contrastive}. 

The proposed framework further integrates visual and textual representations to capture the rich semantic characteristics of fashion products. By combining multimodal representation learning with explicit indirect relational modeling and contrastive alignment, APCL provides a flexible and extensible solution for personalized fashion matching. 
To validate the effectiveness of the proposed approach, we conduct extensive experiments on two widely used benchmark datasets. Experimental results demonstrate that APCL consistently outperforms representative baseline methods across multiple evaluation metrics. Additional analyses show that the proposed framework is particularly effective under sparse-data and cold-start conditions, highlighting the value of explicitly modeling indirect relational structures for recommendation tasks.

The main contributions of this work are summarized as follows:
\begin{itemize}
\item We introduce an \textbf{explicit indirect relational learning framework} for personalized fashion matching, which models indirect user-item and item-item relationships as dedicated learning views rather than relying solely on implicit graph propagation or co-occurrence statistics.

\item We design an \textbf{adaptive preference modeling architecture} that decomposes recommendation into personalization and compatibility dimensions and enriches both through indirect relational representations within a unified framework.

\item We develop a \textbf{functional-view contrastive learning strategy} that aligns direct and indirect relational representations within personalization and compatibility spaces, improving representation consistency and robustness under sparse interaction settings.

\item We conduct comprehensive experiments on benchmark datasets and demonstrate the effectiveness of the proposed framework, particularly in sparse-data and cold-start scenarios.
\end{itemize}

The remainder of the paper is organized as follows. \Cref{Sec:Related_work} reviews related research studies. \Cref{Sec:Method} presents the proposed APCL framework.  \Cref{Sec:experiment} reports experimental evaluations over two public datasets. Finally,~\Cref{sec:concl} concludes the paper and discusses future research directions.

\section{Related Work}
\label{Sec:Related_work}
\subsection{Personalized Fashion Complementary Recommendation}
Personalized fashion recommendation differs from traditional recommendation tasks in that it requires simultaneously modeling user-specific preferences and compatibility relationships among multiple fashion items. Rather than recommending products independently, the objective is to identify items that not only match a user's preferences but also form aesthetically coherent outfit combinations~\cite{deldjoo2023review,GPBPR,CP, ding2026b2}.

Early studies primarily relied on \textbf{collaborative filtering (CF)} techniques~\cite{weijun2014collaborative,cui2020personalized,ebesu2018collaborative,liu2023megcf}, which infer latent preferences from historical user--item interactions. Although effective when sufficient interaction data are available, these methods are often susceptible to severe performance degradation under sparse settings, particularly in fashion domains characterized by rapid product turnover and limited user feedback.

To mitigate sparsity, \textbf{content-based recommendation} approaches~\cite{de2015content,d2024zero} incorporate auxiliary information such as visual attributes, textual descriptions, and category metadata. With advances in computer vision and natural language processing, these approaches have evolved into multimodal representation learning frameworks that integrate visual and textual features to capture semantic characteristics and compatibility relationships among fashion products~\cite{kang2019complete,ma2024personalized, sanny2026medal}. While effective for compatibility modeling, these approaches often provide limited capability for modeling personalized preference signals embedded within interaction data.

More recently, \textbf{hybrid recommendation frameworks} have emerged, combining interaction data with multimodal product information in end-to-end learning architectures~\cite{9428602,liao2023recommendation,sun2025multi}. Representative examples include BPR-based extensions that incorporate visual and textual features to model user preferences~\cite{BPR,vbpr,GPBPR}, sequential recommendation approaches that exploit temporal interaction dynamics~\cite{hui2022personalized,li2020hierarchical,hou2019explainable}, and graph-based models that capture higher-order dependencies among users and fashion items~\cite{wang2019neural,he2020lightgcn,cui2019dressing, cmfn}. These developments reflect a broader trend toward jointly modeling personalization and compatibility within unified recommendation frameworks.

Despite these advances, most existing methods primarily focus on improving representation learning while treating indirect relational signals as implicit outcomes of graph propagation, interaction aggregation, or compatibility learning. Explicitly modeling and disentangling indirect preference and compatibility relationships remains relatively underexplored.

\subsection{Explicit and Implicit Modeling of Indirect Relationships}
Modeling indirect relationships has become an important direction in recommendation research, particularly for addressing data sparsity and cold-start problems. Existing approaches can generally be categorized into implicit and explicit relational modeling paradigms.

\textbf{Implicit relational modeling} is predominantly represented by graph-based recommendation methods. Approaches like NGCF \cite{wang2019neural}, GraphSAGE-based recommendation \cite{ying2018graph}, and LightGCN~\cite{he2020lightgcn} capture higher-order user--item relationships through iterative message passing on interaction graphs. By propagating information across multiple hops, these methods enable users and items to influence one another beyond direct interactions. Although effective, different relational signals are typically aggregated into a shared latent representation space, making it difficult to explicitly distinguish between the semantic roles of different relationships.

A related line of research estimates indirect relationships through \textbf{co-occurrence statistics}~\cite{itemcf,chen2021conet}. These approaches infer associations among users or items based on shared interaction patterns and are computationally efficient. However, because they primarily rely on frequency-based statistics, the resulting relationships often reflect coarse correlations and provide limited capacity for adaptive weighting or semantic interpretation.

More recently, researchers have proposed context-aware, higher-order, and graph-enhanced recommendation frameworks that attempt to exploit richer relational structures~\cite{liu2024unifying,he2023cross,wang2023cross}. Nevertheless, \textbf{indirect relationships} are generally treated as latent structures embedded within existing interaction graphs or neighborhood aggregations rather than as dedicated learning objectives.

In contrast to these approaches, we adopt an \textbf{explicit indirect relational learning} perspective. Instead of relying on graph propagation or static co-occurrence statistics to implicitly encode indirect relationships, the proposed framework explicitly constructs indirect user--item and item--item relationships through correlation-guided sampling and represents them as dedicated learning views. This formulation enables indirect relational signals to be selectively incorporated, adaptively aggregated, and independently analyzed within recommendation modeling.

\subsection{Contrastive Learning for Recommendation}
\textbf{Contrastive learning (CL)} has recently emerged as a powerful paradigm for representation learning in recommendation systems~\cite{guo2025knowledge,jing2023contrastive,wei2021contrastive}. By constructing positive and negative sample pairs and optimizing similarity objectives in latent spaces, CL enables recommendation models to learn robust and discriminative representations under sparse supervision. As a result, CL has become particularly attractive for addressing sparsity and cold-start challenges.

One major line of work focuses on \textbf{interaction-level contrastive learning}, where positive and negative pairs are derived from user--item interactions to regularize representation learning~\cite{wei2021contrastive,guan2019deep,xiao2025nfgcl}. Another line explores \textbf{multi-view contrastive learning}, where different views generated from graph perturbations, side information, or modality-specific representations are aligned within a common embedding space~\cite{wu2022multi,zheng2021multi,zou2022multilevel,ma2022crosscbr,wang2023cross}. In multimodal recommendation, CL is frequently employed to align visual and textual representations or to enhance cross-modal interaction learning~\cite{he2023cross,xu2024collaborative}.

Despite their effectiveness, most existing contrastive recommendation methods define contrastive objectives across modalities, graph augmentations, or alternative data views. Consequently, they largely focus on improving representation consistency without explicitly considering the semantic roles of different relational signals. In particular, direct and indirect relationships are rarely modeled as distinct learning entities within contrastive frameworks.

To address this limitation, we introduce a \textbf{functional-view contrastive learning} paradigm that aligns direct and indirect representations within both personalization and compatibility spaces. Rather than contrasting modalities or graph augmentations, our framework contrasts relational functions. By enforcing consistency between direct and indirect relational views, the proposed approach leverages indirect signals as auxiliary supervision while preserving the semantic structure of personalized fashion matching.

\textit{Overall, existing recommendation approaches capture indirect relational information either implicitly through graph propagation or statistically through co-occurrence patterns. In contrast, this work explicitly constructs indirect user--item and item--item relationships and integrates them through functional-view contrastive alignment. This combination provides a structured framework for modeling indirect relational signals, offering an alternative perspective on addressing sparsity and multimodal recommendation challenges.}

\begin{table}[t]
\caption{List of Key Symbols}
\vspace{-10px}
\begin{center}
\resizebox{\textwidth}{!}{
\begin{tabular}{ll}
\toprule
Symbol  & Description  \\
\midrule
$\mathcal{U}$ &The set of users $\mathcal{U}=\left\{u_1, u_2, \ldots, u_{|\mathcal{U}|}\right\}$.\\
$\mathcal{G}$ &The set of given products $\mathcal{G}=\left\{g_1, g_2, \ldots, g_{|\mathcal{G}|}\right\}$.\\
$\mathcal{R}$ &The set of recommended matching products $\mathcal{R}=\left\{r_1, r_2, \ldots, r_{|\mathcal{R}|}\right\}$.\\
$\mathbf{\bar{v}}$ & The normalized visual features  \\
$\mathbf{\bar{w}}$ & The normalized textual features \\
$\mathbf{\bar{e}}$ & The normalized latent embedding  \\
$\mathcal{L}_{\text{BPR}}$ &The BPR loss function   \\
$\mathcal{L}_{\text{CL}}$ &The contrastive learning loss function   \\
$\mathcal{L}$ &The overall loss function   \\
$p^{u,g}_{r}$ & The preference score of user $u$ to product $r$ for matching with product $g$ \\
$\mathcal{F}(\cdot)$ &The prediction function \\ 
\hline 
\end{tabular}}
\end{center}
\label{tab: symbol}
\end{table}

\section{Method}
\label{Sec:Method}
\subsection{Problem Formulation}
We consider the task of personalized fashion matching, where the objective is to recommend a matching product that is simultaneously compatible with a given product and aligned with a user's preference. \Cref{tab: symbol} defines the notations being used in the personalized fashion matching task.

\textbf{Interaction data.}  Let $\mathcal{U}$ denotes the set of users, $\mathcal{G}$ the set of given fashion products, and $\mathcal{R}$ the set of candidate matching products. 
An outfit is defined as a combination of a given product $g \in \mathcal{G}$ and a recommended matching product $r \in \mathcal{R}$, denoted as $(g, r)$, representing item--item relations. Each user $u \in \mathcal{U}$ engages with several of these pairs, resulting in a set of \textbf{triplets} $<u,g,r>$ that record \textbf{user interactions}. 

\begin{figure*}[t]
  \centering
  \includegraphics[width=\textwidth]{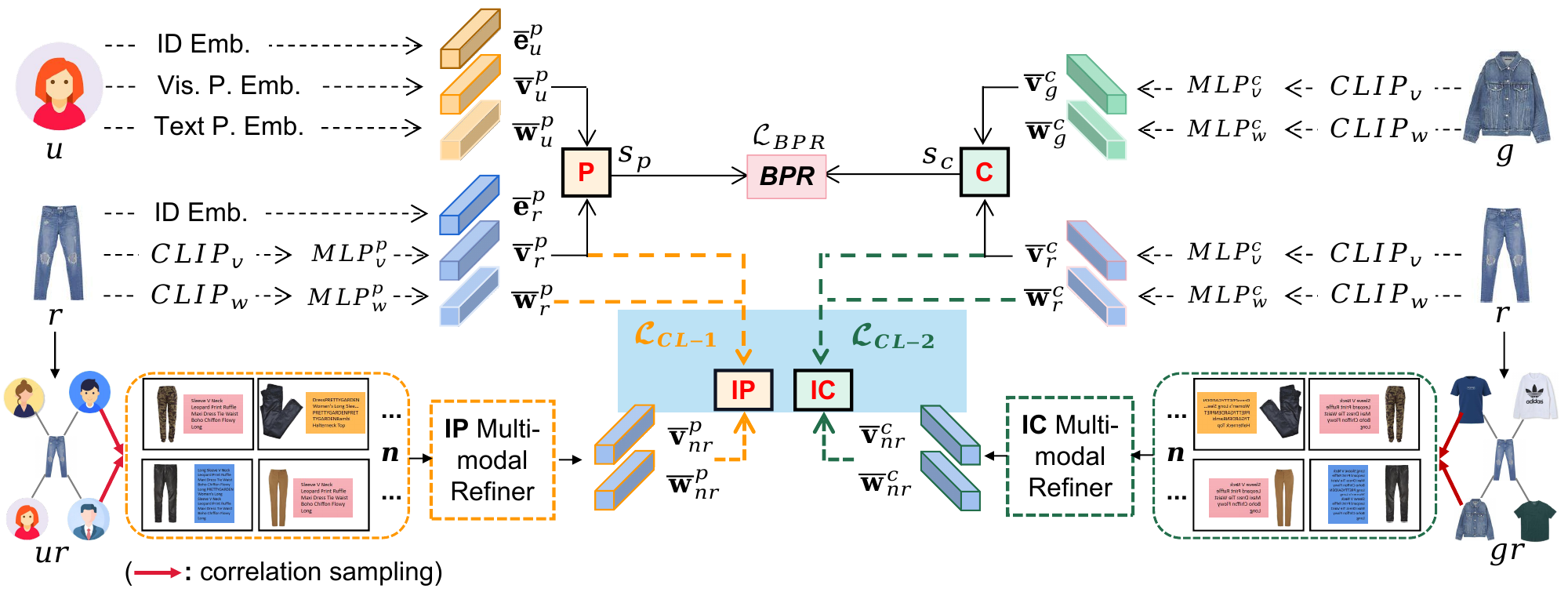}
    \caption{The proposed APCL scheme has four main components: Personal preference module (P); product Compatibility module (C); Indirect Personal preference module (IP); and Indirect product Compatibility module (IC), along with correlation sampling strategy within IP and IC.}
  \label{fig:method}
\end{figure*}

{\textbf{Multi-modal data.}} 
To better characterize the clothing products in $\mathcal{G}$ and $\mathcal{R}$, both visual and textual modalities of features are leveraged and extracted using the pre-trained CLIP model \cite{clip} in this study. For a given product/item $g$, we obtain a visual feature vector $\mathbf{v}_g \in \mathbb{R}^{d_v}$ and a textual feature vector $\mathbf{w}_g \in \mathbb{R}^{d_w}$, where $d_v$ and $d_w$ denote the dimensions of the visual and textual representations, respectively. Similarly, each recommended product/item $r$ is associated with visual and textual features $\mathbf{v}_r$ and $\mathbf{w}_r$.
For each user $u$, a learnable latent embedding assigned with ID ($\mathbf{e}_u^p \in \mathbb{R}^d$) is developed, and a learnable visual latent factor ($\mathbf{e}_u^v$) and a textual latent factor  ($\mathbf{e}_u^t$) represent the user's visual and textual preferences for the recommended product \cite{vbpr, GPBPR}. In this way, a cold-start user embedding therefore can also be derived.
In addition, each recommended product $r$ is also associated with a latent factor $\mathbf{e}_r \in \mathbb{R}^d$. The user’s preference score for product $r$ is modeled by evaluating the similarity between their normalized latent embeddings, $\mathbf{\bar{e}}_u$ and $\mathbf{\bar{e}}_r$, combined with visual and textual bias terms derived from the multi-modal features. 
This approach allows integration of collaborative signals and multi-modal content information to form a unified representation.

{\textbf{Problem statement.}} The goal of personalized fashion complementary recommendation is to predict preference scores for any triplet combination of  $<u, g, r>$: 
\begin{equation}
p^{u,g}_{r} = \mathcal{F}(u, g, r | \Theta),
\label{eq:overall_obj}
\end{equation}
where $\Theta$ are the parameters of the recommendation model $\mathcal{F}$. This score $p^{u,g}_{r}$ signifies the possibility of the recommended product $r$ being preferred by the user $u$ while also being a good match with the given product $g$. The objective is to learn a function $\mathcal{F}$ that scores higher for compatible triplets than incompatible ones.

Unlike conventional recommendation tasks that primarily model direct user--item $(u,g)$ interactions, personalized fashion matching additionally requires modeling compatibility $(g,r)$ relationships among fashion products. Furthermore, many useful recommendation signals originate from indirect relationships among users and products, particularly under sparse interaction settings. Therefore, the proposed framework aims to jointly model both direct and indirect relational signals.

\subsection{Overview of the APCL Framework}
The proposed \textbf{A}daptive \textbf{P}reference with \textbf{C}ontrastive \textbf{L}earning (\textbf{APCL}) framework is designed to explicitly model indirect relational signals for personalized fashion matching.

Unlike graph-based recommendation methods that capture higher-order relationships implicitly through graph propagation~\cite{wang2019neural,he2020lightgcn}, APCL constructs indirect user--item and item--item relationships as \textit{explicit} learning components and organizes them into dedicated relational views.

As illustrated in~\Cref{fig:method}, the framework consists of four modules: (1) Personal Preference Modeling (\textbf{P}), (2) Product Compatibility Modeling (\textbf{C}), (3) Indirect Personal Preference Modeling (\textbf{IP}), and (4) Indirect Compatibility Modeling (\textbf{IC}).
The P and C modules model direct relational signals obtained from observed interactions, whereas the IP and IC modules explicitly capture indirect relational signals derived from correlated users and products. This decomposition enables direct and indirect relationships to be modeled separately while preserving their complementary roles in recommendation.

In this study, a functional-view contrastive learning mechanism is introduced to align direct and indirect representations within both personalization and compatibility spaces.
Specifically, the representation ($s_c$) from multi-modal product compatibility learning view in the compatibility space captures the degree of similarity between a target recommended product $r$ and a given product $g$ in terms of different content characteristics; the representation ($s_p$) from the personal preference leaning view in the personalization space captures the relevant features of the target recommended product and users' personal preferences.  
The final prediction, w.r.t.~\cref{eq:overall_obj}, is then obtained by integrating the outputs from preference and compatibility views:
\begin{equation}
p^{u,g}_{r} = s_p + s_c.
\label{eq.all}
\end{equation}
As \cref{eq.all} infers, the APCL framework departs from traditional personalized complementary recommendation methods, such as GP-BPR~\cite{GPBPR}, which typically treat personalization and compatibility as mutually exclusive or competing factors. 
In contrast, APCL assumes that these two aspects are not inherently contradictory, and allows them to be modeled jointly without imposing a trade-off between the strength of personalization and the level of compatibility.

\subsection{Direct Relationship Modeling}
The first stage of APCL models direct relationships observed in historical interaction data. Specifically, the framework captures two complementary aspects of recommendation: compatibility relationships among products and preference relationships between users and products.

\subsubsection{Product Compatibility Modeling (\textbf{C})}
\label{subsec:compatibility}
The compatibility module models direct item--item relationships by evaluating how well a candidate product $r$ complements a given product $g$. This branch captures observable compatibility signals derived from multimodal product characteristics.

Previous studies have shown that visual and textual information provide complementary evidence for fashion compatibility modeling~\cite{GPBPR,ding2023computational,ma2024personalized}. Visual features capture style coherence and aesthetic consistency, while textual information provides semantic descriptions of product characteristics~\cite{wang2025multimodal,huang2024multimodal}. By jointly leveraging both modalities, the model learns a compatibility space in which compatible fashion products are represented closer to one another.

In the latent space, compatibility is modeled as the similarity between the encoded representations of the given item $g$ and the target matching item $r$.
For each modality, we adopt two multi-layer perceptrons (MLPs) followed by a batch normalization layer as the feature encoder to obtain either visual or textual representation for each fashion product. Specifically, for compatibility learning, we have,
\begin{equation}
\begin{aligned}
&
\left\{
\begin{aligned}
&\mathbf{\bar{v}}^c_g = \text{Norm}(\text{MLP}^c_v(\mathbf{v}_g)), \\
&\mathbf{\bar{v}}^c_r = \text{Norm}(\text{MLP}^c_v(\mathbf{v}_r)), \\
&s_c^v= \frac{\mathbf{\bar{v}}^c_g \cdot \mathbf{\bar{v}}_r^c}{\|\mathbf{\bar{v}}^c_g\| \|\mathbf{\bar{v}}_r^c\|},
\end{aligned}
\right.
&&
\left\{
\begin{aligned}
&\mathbf{\bar{w}}^c_g = \text{Norm}(\text{MLP}^c_w(\mathbf{w}_g)), \\
&\mathbf{\bar{w}}^c_r = \text{Norm}(\text{MLP}^c_w(\mathbf{w}_r)), \\
&s_c^w= \frac{\mathbf{\bar{w}}^c_g \cdot \mathbf{\bar{w}}_r^c}{\|\mathbf{\bar{w}}^c_g\| \|\mathbf{\bar{w}}_r^c\|},
\end{aligned}
\right.
\end{aligned}
\label{eq.scw_and_scv}
\end{equation}
where $\mathbf{\bar{v}}^c_g$ and $\mathbf{\bar{v}}^c_r$ are visual latent representation results of the given product $g$ and the recommended matching product $r$ obtained from the visual compatibility encoder, respectively. The cosine similarity ($s_c^v$) is used to evaluate the similarity between the two products in the learned visual compatibility space. Since cosine similarity is scale-invariant, meaning it is unaffected by the magnitude of the vectors and is commonly used in the application of recommendation systems. In~\cref{eq.scw_and_scv}, $\mathbf{w}$ denotes the results from the textual compatibility encoder. The results of the visual and textual compatibility are combined linearly to provide the comprehensive multi-modal compatibility score ($s_c$):
\begin{equation}
s_c=\pi \cdot s_c^v +(1-\pi) \cdot s_c^w, 
\label{eq.sc}
\end{equation}
where $\pi$ is the trade-off parameter to control the contribution of each modality.

\subsubsection{Personal Preference Modeling (\textbf{P})}
The personal preference module models direct user--item relationships based on observed interaction histories. Unlike the compatibility branch, which focuses on relationships among products, this module aims to capture personalized preference signals associated with individual users.

Following BPR-based personalized recommendation frameworks~\cite{BPR,GPBPR,vbpr}, each user and item is represented through latent embeddings. Visual and textual preference factors are further introduced to capture modality-specific user interests, enabling the framework to preserve personalization information under sparse interactions and cold-start settings. The prediction of users' personal preferences, $s_u$, based on feature-level user-item interactions, is derived as follows:
\begin{equation}
s_u= \frac{\mathbf{\bar{e}}_u^p \cdot \mathbf{\bar{e}}_r^p}{\|\mathbf{\bar{e}}_u\| \|\mathbf{\bar{e}}_r\|}+\beta_u+\beta_r+ \alpha.
\label{eq.sur}
\end{equation}
Here we use learnable ID-linked embeddings to explore users potential preference. $\mathbf{\bar{e}}_u^p$ and $\mathbf{\bar{e}}_r^p$ are normalized latent embeddings for user $u$ and target recommended product $r$, respectively. $\beta_u$ is the user bias term, and $\beta_r$ denotes the product bias term. $\alpha$ is the global offset which normalizes the scoring scale.
In addition, we predict the visual and textual preferences by combining the visual and textual preference variables into latent content-based preference components as follows:
\begin{align}
\begin{aligned}
&s_p^v= \frac{\mathbf{\bar{v}}^p_u \cdot \mathbf{\bar{v}}_r^p}{\|\mathbf{\bar{v}}^p_u\| \|\mathbf{\bar{v}}_r^p\|},
\end{aligned}
&
\begin{aligned}
&s_p^w= \frac{\mathbf{\bar{w}}^p_u \cdot \mathbf{\bar{w}}_r^p}{\|\mathbf{\bar{w}}^p_u\| \|\mathbf{\bar{w}}_r^p\|}.
\end{aligned}
\label{eq.sp_vw}
\end{align}
We use $\mathbf{v}^p_u$ and $\mathbf{w}^p_u$ embeddings to help capture latent user preferences in terms of visual and textual factors, where batch normalization is used to obtain $\mathbf{\bar{v}}^p_u$ and $\mathbf{\bar{w}}^p_u$. For latent representation of matching product $r$, we use two multi-layer perceptrons (MLPs) followed by a batch normalization layer for each modality as the feature encoders: 
\begin{align}
\begin{aligned}
&\bar{\mathbf{v}}^p_{r} = \text{Norm}(\text{MLP}_v(\mathbf{v}_{r})),
\end{aligned}
&
\begin{aligned}
&\bar{\mathbf{w}}^p_{r} = \text{Norm}(\text{MLP}_w(\mathbf{w}_{r})). 
\end{aligned}
\label{eq: mlp}
\end{align}

The total user personal preference modeling is thus obtained by:
\begin{equation}
s_p=s_u+\pi \cdot s_p^v+ (1-\pi) \cdot s_p^w,
\label{eq: sp}
\end{equation}
where $\pi$ is the trade-off parameter to control the contribution of the visual modality and textual modality.

\subsection{Explicit Indirect Relational Learning}
A central idea of APCL is explicit indirect relational learning. We argue that many valuable recommendation signals are not directly observable from individual interactions but emerge through relationships among correlated users and products.

Existing approaches typically capture such signals implicitly through graph propagation~\cite{wang2019neural,he2020lightgcn} or co-occurrence statistics~\cite{itemcf,chen2021conet}. In contrast, APCL explicitly constructs indirect user--item and item--item relationships and models them as dedicated learning views. This design provides greater interpretability and control over the role of indirect relational information during recommendation.


\begin{figure}[!t]
  \centering
  \setlength{\abovecaptionskip}{0.2cm}
  \includegraphics[width=.92\linewidth]{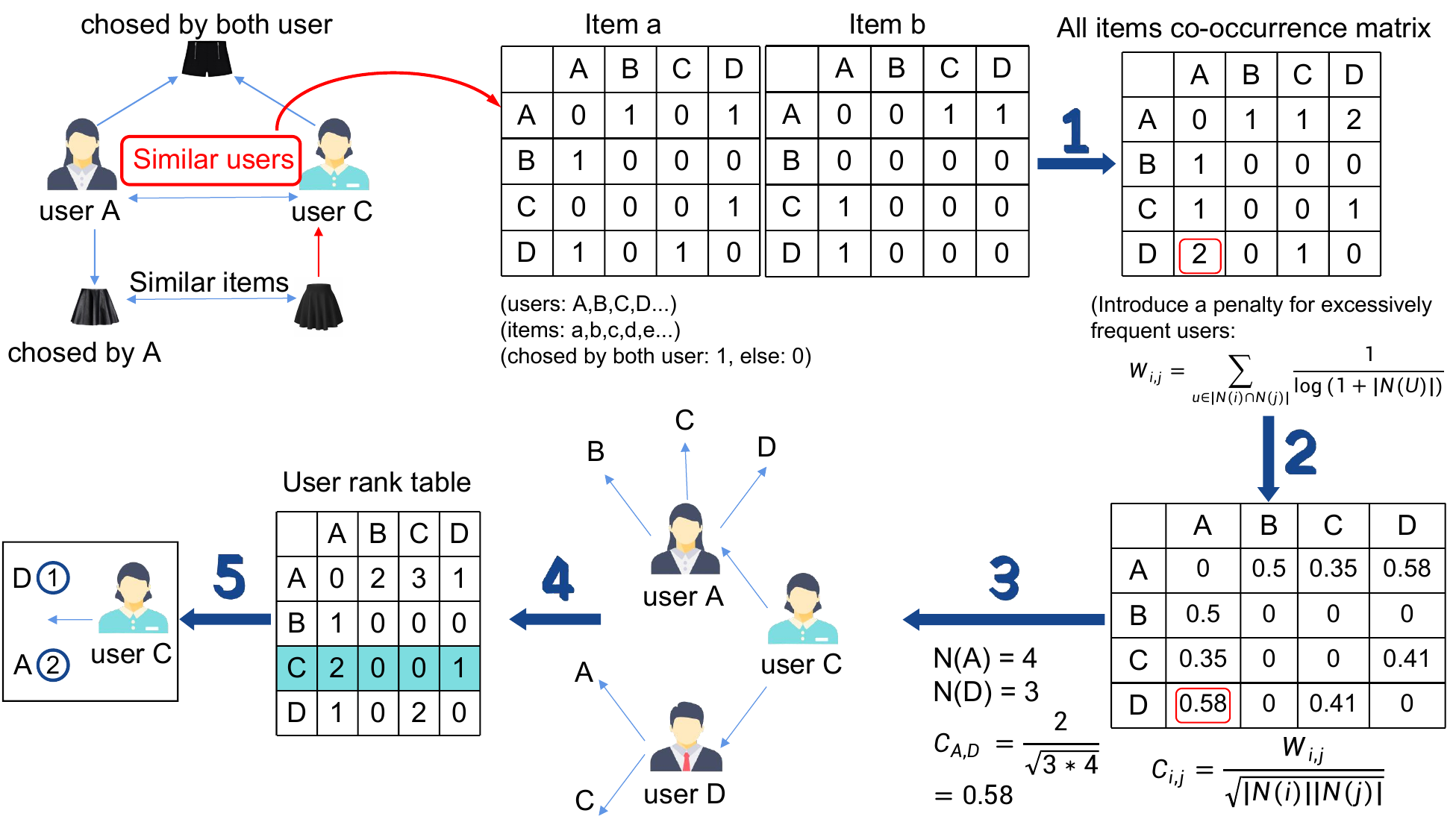}
  \caption{Simplified example of correlation sampling strategy from IP module.}
  \label{fig: usercf}
\end{figure}

\subsubsection{Indirect Personal Preference Modeling (\textbf{IP})}
The IP module explicitly models indirect user--item relationships. 

Given a target user, the module identifies correlated users with similar interaction patterns and utilizes their historical preferences as auxiliary evidence. Rather than treating these relationships as implicit graph connections, APCL constructs them explicitly through a correlation-guided sampling mechanism. The resulting relational view provides additional preference signals that complement the direct interactions captured by the personal preference module.

Specifically, we identify `similar users' who have interacted with the same target matching products before (i.e., $ur$ in~\Cref{fig:method}) and aggregate their interaction patterns to enrich the target user’s representation.
This aggregation is performed using an attention-based mechanism, which assigns higher weights to more relevant signals.

By computing the similarity between the target matching product and positive products selected by similar users, we can identify additional products that share similar features or attributes to the target matching product. Positive products here refer to products that have been chosen and positively evaluated by the users. These products likely reflect similar preferences of users. 

\begin{figure}[t!]
  \centering
  \includegraphics[width=\linewidth]{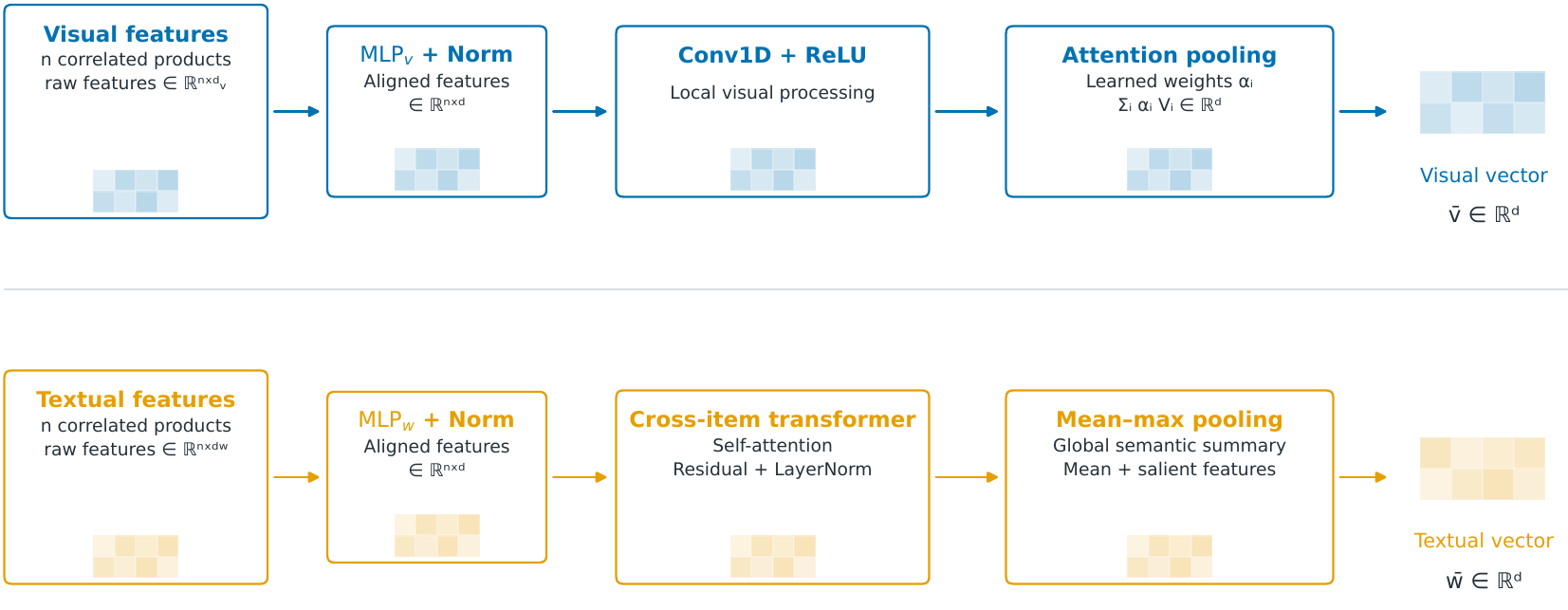}
  \vspace{-15px}
  \caption{Implementation in Multi-Modal Refiner.}
  \vspace{-10px}
  \label{fig: att}
\end{figure}

For implementation details, we first filter users share similar interest through a \textbf{correlation sampling strategy} (see~\Cref{fig: usercf}), then aggregate the $n$ positive matching products chosen by those users. 
After that, we apply an attentive \textbf{multimodal latent refiner} (see ~\Cref{fig: att}) to generate a latent preference representation connected to the target matching product for each modality.

\paragraph{\textbf{Correlation Sampling Strategy}}\label{sec:corr_sample}
The purpose of correlation sampling is to convert sparse interaction records into explicit relational structures. Instead of propagating information across a global graph, the proposed strategy directly identifies correlated entities and constructs local relational neighborhoods for representation learning.

The implementation details are explained as follows, and here we set the $ur$ list selection for IP modeling as an example:

\begin{itemize}
\item {\verb|step 1|}: For each item in the interaction matrix, if both user $A$ and user $B$ have interacted with that product, we record the interaction as 1; otherwise, we record it as 0. This process is repeated for all items in the interaction matrix until all interactions between users and items have been recorded. By summing the matrices, we get a user-user co-occurrence ($W_{i, j}$) matrix (cumulative summation) based on each positive recommended product. 
\item {\verb|step 2|}: Calculate user similarity score ($C_{i, j}$) matrix as follows:
\begin{equation}
C_{i, j}=\frac{W_{i, j}}{\sqrt{|N(i)||N(j)|}}
\end{equation}
where $N(i)$ and $N(j)$ represent the number of occurrences in the whole training set for users $i$ and $j$, respectively.
\item {\verb|step 3|}: List similar users for each user.
\item {\verb|step 4|}: Rank similar users by similarity score ($C_{i, j}$).
\item {\verb|step 5|}: Find top $K$ the most similar users for each target user, and for each similar user, use cross attention to select $n$ related matching products ($ur$) from content-based learning manner.
\end{itemize}

\paragraph{\textbf{Multi-modal Latent Refiner}}\label{sec: Refiner}
After correlation sampling, a set of correlated products are obtained. These products form an explicit indirect relational view associated with the target user. To transform this view into informative latent representations, we introduce a multimodal latent refiner that aggregates visual and textual information while preserving modality-specific characteristics.

Given $n$ pre-filtered items with strong correlations after correlation sampling, our framework processes visual and textual features through shared projections followed by modality-specific refinement (\Cref{fig: att}). As $\mathbf{v}_{ur} \in \mathbb{R}^{d_v}$ and $\mathbf{w}_{ur} \in \mathbb{R}^{d_w}$ denote the $n$ raw visual and textual features respectively. We first project them into aligned spaces like \cref{eq: mlp}:
\begin{equation}
\begin{aligned}
\mathbf{v}^p_{ur} &= \text{Norm}(\text{MLP}_v(\mathbf{v}_{ur})) \in \mathbb{R}^{n \times d} \\
\mathbf{w}^p_{ur} &= \text{Norm}(\text{MLP}_w(\mathbf{w}_{ur})) \in \mathbb{R}^{n \times d}
\end{aligned}
\end{equation}
After the projection, we design a \textbf{dual-pathway} latent refiner, which addresses fundamental differences in visual and textual data structures through distinct processing strategies. 

As visual features exhibit strong \textit{spatial locality} and \textit{translation invariance}, making convolutional operations ideal for pattern extraction. For visual features, we process them using 1D convolutions along the spatial dimension, equivalent to treating each feature dimension as a channel, along with the ReLU activation. To obtain the unified visual representation $\bar{\mathbf{v}}_{ur}^p \in \mathbb{R}^d$:
\begin{equation}
\bar{\mathbf{v}}_{ur}^p = \text{AttnPool}(\mathbf{V}^{ur}) = \sum_{i=1}^n \alpha_i \mathbf{V}^{ur}_i
\end{equation}
The attention weights $\alpha_i$ are computed as:
\begin{equation}
\alpha_i = \text{softmax}(\mathbf{w}^T \tanh(\mathbf{W}_v \mathbf{V}^{ur}_i))
\end{equation}
where $\mathbf{W}_v \in \mathbb{R}^{h \times d}$ and $\mathbf{w} \in \mathbb{R}^h$ are learnable parameters.

For textual feature path, as textual semantics require \textit{long-range dependency modeling}, such as semantic association of product descriptions. And many leading works have demonstrated the superiority of transformer for processing textual information.
The adopted cross-item transformer architecture provides multi-head self-attention with $h=2$ heads, along with Layer normalization and residual connections. 
Then the unified textual representation $\bar{\mathbf{w}}_{nr}^p \in \mathbb{R}^d$ is obtained via the \textit{Mean-Max Pooling} following transformer. As Mean pooling captures overall semantic information and Max pooling preserves salient features, which provides robust aggregation.

\subsubsection{Indirect Product Compatibility Modeling (\textbf{IC})}
The IC module follows the same explicit relational learning principle as the IP module, but focuses on indirect item--item relationships. Specifically, products exhibiting similar compatibility patterns are identified through correlation sampling (discussed in~\Cref{sec:corr_sample}). These correlated products are then aggregated to enrich compatibility representations beyond direct item--item observations. Consequently, latent compatibility cues ($\bar{\mathbf{v}}_{nr}^c$, and $\bar{\mathbf{w}}_{nr}^c$) that are difficult to capture from direct interactions alone can be incorporated into recommendation.

\subsection{Optimization}
\subsubsection{BPR Pairwise Learning}
With triplet interaction data, we employ the widely used BPR to recommend, for a given user and a given product, suitable matching product over a pair of positive and negative samples, by calculating the preference score (\cref{eq:overall_obj}) as $p^{u,g}_{r_+}$ and $p^{u,g}_{r_-}$, respectively. The BPR pairwise loss, $\mathcal{L}_{\text{BPR}}$, on the entire training set $\mathcal{D}$ is determined as follows: 
\begin{equation}
\mathcal{L}_{\text{BPR}}=\sum_{\mathcal{D}}\left[-\ln \left(\sigma\left(p_{r_+}^{u,g}-p_{r_-}^{u,g}\right)\right)\right]+\frac{\lambda}{2}\left\|\Theta_F\right\|^2,
\label{eq.bprloss}
\end{equation}
where $\mathcal{D} = \left\{(u, g, r_+, r_-) \mid u \in \mathcal{U} \wedge g \in \mathcal{G} \wedge (r_+, r_-) \in \mathcal{R} \right\}.$ ${\lambda}$ is a hyper-parameter and ${\Theta_F}$ represents the parameter set of the model, and $\sigma$ is the Sigmoid function.

\subsubsection{Functional-View Contrastive Learning}
The explicit relational views introduced by IP and IC create complementary representations of the same recommendation entities. To encourage consistency between direct and indirect relational views, APCL introduces a functional-view contrastive learning mechanism. Unlike conventional contrastive learning approaches that primarily align modalities, graph augmentations, or perturbed views~\cite{wu2022multi,zou2022multilevel,ma2022crosscbr}, the proposed framework aligns representations across relational functions.

The overall CL module contains two components:
cross P-IP view contrastive learning and cross C-IC view contrastive learning.

\textbf{P-IP contrastive learning:}
We establish our contrastive loss by alimenting the latent target matching product representations obtained through personal preference modeling module ($\mathbf{\bar{v}}^p_r$/$\mathbf{\bar{w}}^p_r$) and indirect personal preference modeling module ($\mathbf{\bar{v}}^p_{nr}$/$\mathbf{\bar{w}}^p_{nr}$) by \cref{eq.sp_vw,eq.scw_and_scv}. The final  contrastive learning for both visual and textual modalities is defined as follows: 
\begin{equation}
\begin{aligned}
&\mathcal{L}_{\text{CL-1}} = \mathcal{L}_{\text{CL-1}}^v + \mathcal{L}_{\text{CL-1}}^w,\\
&\text{where } \mathcal{L}_{\text{CL-1}}^v=\sum_{\mathcal{V}}\left[-\log \frac{e^{ \left(\left(\mathbf{\bar{v}}^p_{r_+} \cdot \mathbf{\bar{v}}^{p}_{nr_+}\right) / t\right)}}{e^{ \left(\left(\mathbf{\bar{v}}^p_{r_+} \cdot \mathbf{\bar{v}}^{p}_{nr_+}\right) / t\right)}+ \sum e^{ \left(\left(\mathbf{\bar{v}}^p_{r_+} \cdot \mathbf{\bar{v}}^p_{r_-}\right) / t\right)}}\right],\\
&\mathcal{L}_{\text{CL-1}}^w=\sum_{\mathcal{W}}\left[-\log \frac{e^{ \left(\left(\mathbf{\bar{w}}^p_{r_+} \cdot \mathbf{\bar{w}}^{p}_{nr_+}\right) / t\right)}}{e^{ \left(\left(\mathbf{\bar{w}}^p_{r_+} \cdot \mathbf{\bar{w}}^{p}_{nr_+}\right) / t\right)}+ \sum e^{ \left(\left(\mathbf{\bar{w}}^p_{r_+} \cdot \mathbf{\bar{w}}^p_{r_-}\right) / t\right)}}\right].
\label{eq_cl1}
\end{aligned}
\end{equation}
where $t$ is the temperature hyper-parameter.
Similarly, the subscript ${r_+}$ and ${r_+}$ represent the pair of positive and negative samples. 

The objective of $\mathcal{L}_{CL-1}$ is to align direct preference representations with their corresponding indirect preference representations while separating them from unrelated negative samples. Consequently, indirect user-level relational signals become an additional source of supervision that enhances preference learning under sparse interaction conditions.

The temperature parameter $t$ enables the exploration of different combinations of personalization and indirect personalization features, thus increasing the diversity of recommendations. At the same time, this loss function allows the model to focus by comparing the similarity of positive and negative samples.

\textbf{C-IC contrastive learning:}
Similarly to the the P-IP contrastive learning, the C-IC view contrastive learning compares the target matching product latent representations derived from the Product Compatibility module: ($\mathbf{\bar{v}}^{c}_{r}$/$\mathbf{\bar{w}}^{c}_{r}$) and adaptive Indirect Compatibility module ($\mathbf{\bar{v}}^{c}_{nr}$/$\mathbf{\bar{w}}^{c}_{nr}$) as follows:
\begin{equation}
\begin{aligned}
&\mathcal{L}_{\text{CL-2}} = \mathcal{L}_{\text{CL-2}}^v + \mathcal{L}_{\text{CL-2}}^w, \\
&\text{where }\mathcal{L}_{\text{CL-2}}^v=\sum_{\mathcal{V}}\left[-\log \frac{e^{ \left(\left(\mathbf{\bar{v}}^{c}_{r+} \cdot \mathbf{\bar{v}}^{c}_{nr+}\right) / t\right)}}{e^{ \left(\left(\mathbf{\bar{v}}^{c}_{r+} \cdot \mathbf{\bar{v}}^{c}_{nr+}\right) / t\right)}+ \sum e^{ \left(\left(\mathbf{\bar{v}}^{c}_{r+} \cdot \mathbf{\bar{v}}^{c}_{r-}\right) / t\right)}}\right],\\
&\mathcal{L}_{\text{CL-2}}^w=\sum_{\mathcal{W}}\left[-\log \frac{e^{ \left(\left(\mathbf{\bar{w}}^{c}_{r+} \cdot \mathbf{\bar{w}}^{c}_{nr+}\right) / t\right)}}{e^{\left(\left(\mathbf{\bar{w}}^{c}_{r+} \cdot \mathbf{\bar{w}}^{c}_{nr+}\right) / t\right)}+ \sum e^{ \left(\left(\mathbf{\bar{w}}^{c}_{r+} \cdot \mathbf{\bar{w}}^{c}_{r-}\right) / t\right)}}\right].
\label{eq_cl2}
\end{aligned}
\end{equation}
Similarly, $\mathcal{L}_{CL-2}$ aligns direct and indirect compatibility representations. This objective encourages compatibility cues captured through indirect item-item relationships to remain consistent with those learned from direct observations, resulting in more robust compatibility modeling.

\subsubsection{Joint Optimization}
The overall optimization objective integrates direct relationship modeling, explicit indirect relational learning, and functional-view representation alignment within a unified learning framework.

The overall objective of APCL model is defined as follows: 
\begin{equation}
\mathcal{L} = \gamma_1 \cdot \mathcal{L}_{\text{BPR}}+ \gamma_2 \cdot (\mathcal{L}_{\text{CL-1}}+ \mathcal{L}_{\text{CL-2}}), \\
\label{eq.allloss}
\end{equation}
where $\gamma_1$ and $\gamma_2$ are the weights for the BPR, the cross P-C view contrastive learning ($\mathcal{L}_{\text{CL-1}}$), and the cross IP-IC view contrastive learning $\mathcal{L}_{\text{CL-2}}$) objectives, respectively. The aim of BPR is to perform personalized ranking in the recommender system. It trains the model by optimizing the relative order of products, which the user actually likes (positive samples) and which the user may not like (negative samples). Although the other two contrastive losses are based on Information Noise-Contrastive Estimation (InfoNCE) \cite{infoNCE}, which represents a contrast loss for unsupervised or semi-supervised learning, it trains the model by optimizing the distance between positive samples (similar data pairs) and negative samples (dissimilar data pairs). The weighted linear combination enables a joint optimization for our indirect connection enriched network for personalized fashion complementary recommendations. 

\subsection{Discussion}
The key contribution of APCL lies in its explicit indirect relational learning paradigm. 

Existing recommendation models typically capture indirect relationships implicitly through graph propagation or statistical co-occurrence. In contrast, APCL explicitly constructs indirect user--item and item--item relationships and models them as dedicated relational views. This formulation enables indirect signals to be separately represented, selectively aggregated, and directly optimized. 

Furthermore, the proposed functional-view contrastive learning mechanism aligns direct and indirect relational representations within personalization and compatibility spaces, resulting in more robust representation learning and improved recommendation performance under sparse interaction settings.
\subsubsection{Data Flow Overview}

To provide a clear overview of the overall architecture and computational flow of the APCL model, we present the forward data flow pseudocode in~\Cref{alg:apcl_forward}. 

\begin{algorithm}[t!]
\caption{Forward Data Flow of APCL Model}
\label{alg:apcl_forward}
\begin{algorithmic}[1]
\REQUIRE Triplet $(u, g, r)$; user set $\mathcal{U}$; item sets $\mathcal{G}, \mathcal{R}$; multimodal features $\mathbf{v}_g, \mathbf{w}_g, \mathbf{v}_r, \mathbf{w}_r$
\ENSURE Prediction score $p_r^{u,g}$

\STATE \textbf{Feature Encoding:}
\STATE $\forall x \in \{g, r\}, m \in \{v, w\}:$
\STATE $\quad \overline{\mathbf{m}}_x^c \gets \text{Norm}(\text{MLP}_m^c(\mathbf{m}_x))$ \COMMENT{Compatibility encoding}
\STATE $\quad \overline{\mathbf{m}}_x^p \gets \text{Norm}(\text{MLP}_m^p(\mathbf{m}_x))$ \COMMENT{Preference encoding}

\STATE \textbf{User embeddings:} $\overline{\mathbf{e}}_u^p$; $\overline{\mathbf{v}}_u^p$; $\overline{\mathbf{w}}_u^p$

\STATE \textbf{Compute personal preference score $s_p$}
\STATE $s_u \leftarrow \cos(\overline{\mathbf{e}}_u^p, \overline{\mathbf{e}}_r^p) + \beta_u + \beta_r + \alpha$
\STATE $s_p^v \leftarrow \cos(\overline{\mathbf{v}}_u^p, \overline{\mathbf{v}}_r^p)$
\STATE $s_p^w \leftarrow \cos(\overline{\mathbf{w}}_u^p, \overline{\mathbf{w}}_r^p)$
\STATE $s_p \leftarrow s_u + \pi \cdot s_p^v + (1 - \pi) \cdot s_p^w$

\STATE \textbf{Compute compatibility score $s_c$}
\STATE $s_c^v \leftarrow \cos(\overline{\mathbf{v}}_g^c, \overline{\mathbf{v}}_r^c)$
\STATE $s_c^w \leftarrow \cos(\overline{\mathbf{w}}_g^c, \overline{\mathbf{w}}_r^c)$
\STATE $s_c \leftarrow \pi \cdot s_c^v + (1 - \pi) \cdot s_c^w$

\STATE \textbf{Adaptive Preference (AP) Module}
\STATE $\mathcal{N}_u \leftarrow \text{TopKSimilarUsers}(u, K)$ \COMMENT{Correlation sampling}
\STATE $\mathcal{I}_u \leftarrow \text{CrossAttention}(\mathcal{N}_u, r)$ \COMMENT{Select $n$ related items}
\STATE $\overline{\mathbf{v}}_{nr}^p, \overline{\mathbf{w}}_{nr}^p \leftarrow \text{MultiModalRefiner}(\mathcal{I}_u)$ \COMMENT{Indirect preference}
\STATE $\overline{\mathbf{v}}_{nr}^c, \overline{\mathbf{w}}_{nr}^c \leftarrow \text{MultiModalRefiner}(\mathcal{I}_u)$ \COMMENT{Indirect compatibility}

\STATE \textbf{Contrastive Loss Computation (during training)}
\STATE $\mathcal{L}_{CL-1} \leftarrow \text{InfoNCE}(\overline{\mathbf{v}}_r^p, \overline{\mathbf{v}}_{nr}^p) + \text{InfoNCE}(\overline{\mathbf{w}}_r^p, \overline{\mathbf{w}}_{nr}^p)$
\STATE $\mathcal{L}_{CL-2} \leftarrow \text{InfoNCE}(\overline{\mathbf{v}}_r^c, \overline{\mathbf{v}}_{nr}^c) + \text{InfoNCE}(\overline{\mathbf{w}}_r^c, \overline{\mathbf{w}}_{nr}^c)$

\STATE \textbf{Final Prediction}
\STATE $p_r^{u,g} \leftarrow s_p + s_c$
\RETURN $p_r^{u,g}$
\end{algorithmic}
\label{alg:apcl_forward}
\end{algorithm}

\subsubsection{Explicit versus Implicit Relational Learning}
Graph-based collaborative filtering methods, such as NGCF~\cite{wang2019neural} and LightGCN~\cite{he2020lightgcn}, capture indirect relationships through iterative propagation over interaction graphs. Although effective, indirect signals remain entangled within the propagation process. 

In contrast, APCL explicitly identifies correlated users and products, constructs dedicated indirect relational views, and aligns these views through contrastive learning. This explicit formulation provides improved interpretability and allows indirect preference and compatibility signals to be individually controlled and analyzed.

\subsubsection{Complexity Analysis}
\label{subsec:complexity}

We analyze the computational and memory costs of the proposed APCL framework. The key variables used in this analysis are summarized in~\Cref{tab:complexity_vars}.

\begin{table}[h]
\centering
\caption{Notations for complexity analysis.}
\label{tab:complexity_vars}
\begin{tabular}{cl}
\hline
\textbf{Symbol} & \textbf{Definition} \\
\hline
$B$ & Batch size \\
$L$ & Maximum number of interactions per user \\
$D$ & Embedding dimension \\
$E$ & Number of attention heads \\
$K$ & Number of similar users/items in AP module \\
$n$ & Number of sampled items in AP module \\
\hline
\end{tabular}
\end{table}

\textbf{Time complexity analysis.}  
The computational cost of APCL mainly arises from multimodal encoding, preference and compatibility scoring, the adaptive preference (AP) module, and contrastive learning. For multimodal encoding, projecting visual and textual features into a latent representation via MLP layers leads to a time complexity of $O(BD^2)$. The personalization and compatibility scoring step requires pairwise similarity computation, which adds $O(BD)$. The AP module introduces two types of operations: correlation sampling with complexity $O(BKL)$, and cross-attention based multi-modal refinement, which incurs $O(BnD^2 + Bn^2D)$. Finally, the InfoNCE loss used in the contrastive learning stage requires similarity evaluation across sampled positive and negative pairs, adding $O(BnD)$. Summing these terms, the overall time complexity per batch can be expressed as $O(BD^2 + BKL + BnD^2 + Bn^2D)$. Since $K$ and $n$ are relatively small compared with $|\mathcal{U}|$ and $|\mathcal{R}|$, the complexity grows linearly with the batch size and embedding dimension, ensuring the scalability of APCL for large-scale datasets.

\textbf{Space complexity analysis.}  
The parameter space of APCL consists of user and item embedding matrices, multimodal encoders, and attention parameters. The embeddings contribute $O(|\mathcal{U}|D + |\mathcal{R}|D)$ space. Each MLP encoder requires $O(D^2)$ parameters, while the cross-attention mechanism in the AP module adds $O(ED^2)$. During forward propagation, intermediate states such as user/item embeddings, sampled item features, and attention weights are cached, incurring an additional cost of $O(BLD + BnD)$. Therefore, the overall space complexity of APCL is $O(|\mathcal{U}|D + |\mathcal{R}|D + ED^2 + BLD + BnD)$.
 
In summary, the time cost of APCL is dominated by $O(BD^2)$ from multimodal encoding and $O(BnD^2)$ from attention-based refinement, while the memory cost is primarily driven by the embedding matrices. Both complexities scale linearly with the batch size and embedding dimension, indicating that APCL achieves a favorable trade-off between expressive modeling power and computational efficiency, making it suitable for deployment in real-world fashion recommendation systems.

\section{Experiment}
\label{Sec:experiment}
To evaluate the effectiveness of explicit indirect relational learning in personalized fashion matching, we conduct comprehensive experiments designed to answer the following research questions:

\textbf{RQ1}: Does APCL outperform representative recommendation methods on personalized fashion matching tasks?

\textbf{RQ2}: What are the respective contributions of direct relational modeling, explicit indirect relational learning, and functional-view contrastive alignment?

\textbf{RQ3}: How robust is the proposed framework with respect to different hyperparameter settings?

\textbf{RQ4}: Does explicit indirect relational learning improve performance under sparse-data, long-tail, and cold-start scenarios?

\subsection{Experimental Setup}
\begin{table}[!t]
\caption{Statistics of datasets used for experimental analysis}
\centering
\renewcommand{\arraystretch}{1.1}
\setlength{\tabcolsep}{8pt}
\begin{tabular}{lrr}
\toprule
Dataset & Polyvore-519 & IQON3000 \\
\midrule
\textbf{User Statistics} &519 & 3,236  \\
\addlinespace
\textbf{Garment Statistics} & & \\
\quad Top garments & 16,887 & 99,655 \\
\quad Bottom garments & 16,220 & 43,086 \\
\quad \textbf{Total garments} & \textbf{33,107} & \textbf{142,737} \\
\addlinespace
\textbf{Sample Statistics} & & \\
\quad Training samples & 39,133 & 170,601 \\
\quad Validation samples & 5,110 & 23,095 \\
\quad Testing samples & 5,197 & 23,095 \\
\quad \textbf{Total samples} & \textbf{49,440} & \textbf{216,791} \\
\bottomrule
\end{tabular}
\label{dataset}
\end{table}

\subsubsection{Datasets} 
We conducted our experiments on two publicly available e-commerce benchmark datasets: \textbf{IQON3000}~\cite{GPBPR} and \textbf{Polyvore}~\cite{polyvore}. Both datasets are widely adopted in the field of personalized fashion recommendation and have served as common benchmarks in numerous prior studies~\cite{GPBPR,PCE,CP}. Detailed statistics including the number of users, items, interactions, and samples of the two datasets are summarized in~\Cref{dataset}.
Following the common setup in prior work, we constructed triplet data samples from user-outfit interactions. Specifically, we extracted pairs of fashion items that belong to the same outfit and are of complementary categories. 
To maintain a controlled evaluation environment, we retain only outfit instances containing exactly one top garment and one bottom garment. Although this simplification does not fully capture the complexity of real-world outfit recommendation involving multiple clothing categories, it enables a focused evaluation of the core personalized fashion matching problem and facilitates fair comparison with previous studies~\cite{GPBPR,PCE,CP}.

\subsubsection{Evaluation Metrics} 
We employed a suite of distinct metrics to thoroughly evaluate the performance of our proposed model: AUC, HR@K, NDCG@K. The AUC metric \cite{AUC}, derived from the Receiver Operating Characteristic (ROC) curve, was utilized to gauge the model's ability to distinguish between positive and negative matching product across the entire test dataset. 
The Hits Ratio at K (HR@K) evaluates the percentage of relevant items that are included within the top K recommendations, serving as an indicator of user satisfaction. The Normalized Discounted Cumulative Gain at K (NDCG@K) is a ranking metric that considers both the relevance of items and their ranking positions. For our analysis, we set K to 10, aligning with common practice in the literature for comparative purposes.
By utilizing this diverse set of evaluation metrics, we demonstrate the proposed model's ability to accurately capture user preferences.

\subsubsection{Baseline models} 
To comprehensively evaluate the effectiveness of the proposed APCL model, we compare it against several representative  baseline models:
\begin{itemize}
\item \textbf{BPR-MF}~\cite{BPR} applies the Matrix Factorization (MF) to capture the latent user-item relations with the Bayesian Personalized Ranking (BPR) algorithm.
\item \textbf{V-BPR}~\cite{vbpr} integrates visual features into the BPR model and captures specific visual preferences of users. 
\item \textbf{T-BPR}~\cite{GPBPR} adopts the same algorithm as V-BPR, replacing the visual features with textural features to leverage the textual information into the BPR modeling. 
\item \textbf{VT-BPR}~\cite{GPBPR} combines V-BPR and T-BPR by further comprehensively characterizing user preferences on the basis of both visual and textual factors.
\item \textbf{GP-BPR}~\cite{GPBPR} is a multi-modal feature-based personalized compatibility modeling method. It effectively models both user preference and product matching relations with a joint BPR framework form. 
\item \textbf{PCE-NET}~\cite{PCE} leverages attention-based compatibility embedding modeling and personal preference modeling to capture fine-grained fashion compatibility features for enhanced personalized recommendation.
\item \textbf{DGSR} \cite{DGSR}:A method for sequential fashion recommendation that models third-order interaction data among users and fashion items.
\item \textbf{TransRec+V} \cite{TransRec}: An visual extended version of TransRec that is a translation-based method for sequential recommendation, learning the latent transition space where item transitions made by different users can be measured with embeddings.
\item \textbf{CP\_TransMatch} \cite{CP} addresses the personalized fashion matching task by leveraging a single-component translation operation to capture third-order user-item interactions and enhancing it with two graph learning modules focused on context and path perspectives.
\item \textbf{HMGL-OCM} \cite{HMGL-OCM} constructs hierarchical graphs at item and modality levels to capture intra-item multimodal interactions and inter-item compatibility. It employs cross-
modality propagation with LSTM to integrate visual, textual, and category features while preserving modal-specific information.
\item \textbf{DR-PCM} \cite{drpcm} uses Homogeneous and heterogeneous graphs, dynamic sampling and adaptive fusion, improves personalized recommendation performance with multi-modalities.
\end{itemize}

The selected baselines represent three major recommendation paradigms:
\begin{enumerate}[label=\alph*)]
    \item direct interaction modeling methods (e.g., BPR-MF, V-BPR, GP-BPR),
    \item graph-enhanced and higher-order relational recommendation methods (e.g., CP-TransMatch, HMGL-OCM, DR-PCM), and
    \item sequential recommendation methods (e.g., DGSR and TransRec+V).
\end{enumerate}
This setup enables us to evaluate whether explicit indirect relational learning provides additional benefits beyond direct interaction modeling, graph propagation, and sequential preference modeling.
\textit{This study focuses on indirect relational learning for personalized fashion compatibility using interaction records and visual/textual features. We therefore compare APCL with non-LLM baselines relevant to compatibility and relational modeling. LLM-assisted methods \cite{cfalr, shi2025integrating, karra2024interarec} introduce additional semantic representations and model components that require a broader experimental design. Their evaluation is beyond the present scope, and comparisons under aligned evaluation protocols remain an important direction for future work.}

\subsubsection{Implementation and training details}
All models were implemented and evaluated in the PyTorch framework and trained with the Adam optimizer~\cite{kinga2015method}. To ensure a fair and reproducible comparison, we applied a uniform tuning and training protocol across all re-implemented baselines and our APCL model. Concretely, we performed a grid search on validation sets with the following ranges: batch size in $[64, 128, 256, 512, 1024]$, learning rate from $10^{-2}$ to $10^{-5}$, weight decay from $10^{-3}$ to $10^{-7}$, and hidden dimension in $[256, 512]$. During training we monitored convergence using early stopping on the validation set. 
To ensure fairness, all baselines were reimplemented under the same preprocessing and evaluation protocol wherever possible. 
For baselines with publicly available code, we used the authors' official implementations. 
For baselines without publicly available code, we implemented the algorithms strictly following the original papers (architectures, loss functions and training protocols). To avoid confounding factors, we standardized data preprocessing, random seeds, optimizer choice, and the hyperparameter search procedure for all re-implemented baselines. All tunable hyperparameters were searched with the same grid described above; the reported value for each model is the best performance obtained on the validation set under this protocol.
For methods whose official code or required auxiliary data were unavailable, we report results from the original publications for reference, and clearly separate these from reproduced results, e.g., 
DGSR \cite{DGSR}, TransRec+V \cite{TransRec}, and HMGL-OCM~\cite{HMGL-OCM}). In particular, HMGL-OCM used additional information beyond textual/visual features and the authors did not provide code, which makes a direct re-implementation non-equivalent, therefore we include their published numbers only for reference.
All results presented in~\Cref{table_all_1,table_all_2} are the best performances achieved after the hyperparameter search described above. MF-BPR was not tuned for content-related hyperparameters since it has no content-based components. By adopting a common optimizer, search ranges, negative sampling strategy and evaluation protocol, we minimize implementation and tuning bias and obtain a reliable empirical comparison of algorithmic effectiveness.

\subsection{Quantitative Evaluations (\textbf{RQ1})}
We compare the proposed APCL method with different baselines on Polyvore dataset and IQON3000 dataset when top and bottom products are respectively specified as the given product ($g$) for recommendation of matching products. 

\begin{table}[h]
\caption{Performance comparison on IQON3000 and Polyvore datasets in terms of setting TOP garment as given product and BOTTOM garment as matching product. Larger numbers indicate better performance ($\uparrow$). \textbf{Bold numbers} indicate the best results, \underline{\textit{italic and underlined}} indicate the second best results.}
\vspace{-10px}
\begin{center}
\small  
\setlength{\tabcolsep}{4mm}{
\begin{threeparttable}
\begin{tabular}{c|cc|cc}
\hline 
\multirow{2}{*}{Methods} & \multicolumn{2}{c|}{IQON3000} &\multicolumn{2}{c}{Polyvore} \\
\cline{2-5}
~ & AUC$\uparrow$ & HR@10$\uparrow$  &\multicolumn{1}{|c}{AUC$\uparrow$} & HR@10$\uparrow$  \\
\hline
BPR-MF \cite{BPR} & 0.8309 &0.7024    & 0.7639  &0.6916 \\ 
T-BPR  \cite{GPBPR}& 0.8316 &0.6361    & 0.7839 &0.6502 \\

V-BPR  \cite{vbpr} & 0.836  &0.6941   & 0.8074  &0.7141\\

VT-BPR  \cite{GPBPR} & 0.8384 & 0.7003& 0.8105  &0.6846 \\

GP-BPR \cite{GPBPR}  & 0.8569 &0.7396    & 0.8232  &0.7121 \\
PCE-NET\cite{PCE} &0.8341   & 0.6399 &0.8235  &0.4208 \\ 
DGSR \cite{DGSR}& 0.8320 &0.6394   & 0.8742 & 0.6645 \\
TransRec+V \cite{TransRec}& 0.8284 & 0.6414  &0.8694 &0.6575 \\
CP-TransMatch \cite{CP} &0.8842 &\underline{\textit{0.8789}}     &0.9001 &\underline{\textit{0.8573}}  \\ 
HMGL-OCM \cite{HMGL-OCM}& \underline{\textit{0.9458}} & --   &\underline{\textit{0.9372}} & -- \\
DR-PCM \cite{drpcm} &0.8879 & 0.6080  & -- & -- \\
\hline 
\textbf{APCL (Ours)} & \textbf{0.9739} &\textbf{0.9509}  &\textbf{0.9832} &\textbf{0.8999} \\
Improvement & 10.1\%$\uparrow$ & 8.2\%$\uparrow$ & 9.2\%$\uparrow$& 5.0\%$\uparrow$\\
\hline
\end{tabular}
\begin{tablenotes}\footnotesize
\item[$\dagger$] Reported results directly from the original publications; re-implementation is not possible due to unavailable of codes or missing proprietary metadata.
\item Moreover, in order to facilitate pair sampled t-tests, the \underline{\textit{second best}} results were highlighted from those cases that re-implementations are possible. 
\end{tablenotes}
\end{threeparttable}
}
\end{center}
\label{table_all_1}
\end{table}

\begin{table}[h]
\caption{Performance comparison on IQON3000 and Polyvore datasets in terms of setting BOTTOM garment as given product and TOP garment as matching product. Larger numbers indicate better performance ($\uparrow$). \textbf{Bold numbers} indicate the best results, \underline{\textit{italic and underlined}} indicate the second best results.}
\vspace{-10px}
\begin{center}
\small
\setlength{\tabcolsep}{0.6mm}{\begin{tabular}{c|ccc|ccc}
\hline 
\multirow{2}{*}{Methods} & \multicolumn{3}{c|}{IQON3000} &\multicolumn{3}{c}{Polyvore} \\
\cline{2-7}
~ & AUC$\uparrow$ & HR@10$\uparrow$ &NDCG@10$\uparrow$  &\multicolumn{1}{|c}{AUC$\uparrow$} & HR@10$\uparrow$ &NDCG@10$\uparrow$ \\
\hline
BPR-MF \cite{BPR} & 0.6849  &0.5598 &0.3776    &0.5724 &0.2598 & 0.1331 \\ 
T-BPR  \cite{GPBPR}& 0.7510 &0.5315 &0.3493   & 0.6250 &\underline{\textit{0.2938}} &0.1638    \\

V-BPR  \cite{vbpr} & 0.7626  &0.5609 &0.3724   & 0.6644  &0.2576 &0.1316  \\

VT-BPR  \cite{GPBPR} & 0.7890 & 0.5341 &0.3576   & 0.6785  &0.2523 &0.1263   \\

GP-BPR \cite{GPBPR}  & 0.7913 &0.6137 &\underline{\textit{0.4663}}  & 0.7075  &0.2500 &0.1253  \\
PCE-NET\cite{PCE} &0.7585   & 0.5111& 0.3239 &0.7072  &0.2222 & 0.1120  \\ 
CP-TransMatch \cite{CP} &\underline{\textit{0.8523}} & \underline{\textit{0.6251}} &0.3756 &\underline{\textit{0.7281}} &0.2859  &\underline{\textit{0.1993}}  \\ 
\hline 
\textbf{APCL (Ours)} & \textbf{0.9016} &\textbf{0.8588} &\textbf{0.7992}  &\textbf{0.7855} &\textbf{0.4880} &\textbf{0.3302}  \\
Improvement & 5.8\%$\uparrow$ & 37.4\%$\uparrow$ & 71.4\%$\uparrow$& 7.9\%$\uparrow$& 4.6\%$\uparrow$& 65.7\%$\uparrow$\\
\hline
\end{tabular}}
\end{center}
\label{table_all_2}
\end{table}

Several important observations can be drawn from \Cref{table_all_1} and~\ref{table_all_2}.

\begin{enumerate}[label=(\arabic*)]
\item APCL consistently achieves the best overall performance across both datasets and recommendation settings. These results suggest that explicitly modeling indirect relational signals provides complementary information beyond that contained in direct user--item and item--item interactions.
\item Models that incorporate multimodal information, including V-BPR, VT-BPR, GP-BPR and PCE-NET, generally outperform purely interaction based approaches. This confirms the importance of visual and textual representations in fashion recommendation.
\item Graph-enhanced approaches such as CP-TransMatch and HMGL-OCM perform substantially better than traditional recommendation models, indicating the importance of exploiting higher-order relational information. However, these methods primarily capture indirect relationships through graph propagation. In contrast, APCL explicitly constructs indirect relational views and therefore provides more direct supervision for learning indirect preference and compatibility signals.
\item The performance gap between APCL and competing methods is particularly noticeable under more sparse recommendation settings, such as recommending matching tops from bottom garments. This observation is consistent with our hypothesis that explicit indirect relational learning is particularly beneficial when direct interaction evidence is limited.
\item The consistent improvements across both datasets demonstrate that the proposed framework generalizes across different interaction distributions and product collections.
\end{enumerate}
In summary, APCL achieves substantial improvements over prior methods. This can be attributed to the joint modeling of direct and indirect relationships together with cross-view alignment, which provides stronger supervision signals under sparse data conditions.

The relatively large improvements can be attributed to the explicit integration of indirect relational signals. Existing methods primarily rely on direct interactions or implicit graph propagation, whereas APCL introduces dedicated indirect relational views and aligns them through contrastive learning. This additional supervision becomes particularly valuable under sparse interaction settings.

\begin{table*}[t]
\caption{Ablation Experiment. -w/o refers to the evaluation results of the models WITHOUT applying according modules.}
\begin{center}
\small
\setlength{\tabcolsep}{1mm}
\begin{tabular}{{ll|c|cccccc|c}}
\toprule[\arrayrulewidth]
\multicolumn{2}{c|}{Dataset} &Metrics ($\uparrow$) & -w/o-V & -w/o-T & -w/o-AP & -w/o-CL & -w/o-P & -w/o-C  & \textbf{APCL}  \\
\midrule
\multirow{8}{*}{\rotatebox{90}{Specified Top}} & \multirow{4}{*}{\rotatebox{90}{Polyvore}}
& AUC &0.8016 &\textit{0.9599} &0.9271 &0.9434 &0.9248 &0.9459 & \textbf{0.9832}\\

~&~&HR@10 &0.4104 &\textit{0.8730} &0.7741 &0.8328 &0.8251 &0.8355 &\textbf{0.8999} \\

~&~&NDCG@10 &0.2354 &\textit{0.7542} &0.6166 &0.6897 &0.7156 & 0.6844 &\textbf{0.8343} \\

~&~&MRR@10 &0.1819 &\textit{0.7167} &0.5670 &0.6447 &0.6808 &0.6363 &\textbf{0.8123} \\

\cmidrule(lr){2-10}
~&\multirow{4}{*}{\rotatebox{90}{IQON}}
&AUC  &0.9541 &0.9296 &0.9528 &\textit{0.9635} &0.6654 &0.9581 &\textbf{0.9739}\\

~&~&HR@10  &0.9244 &0.8780 &0.9128 &\textit{0.9345} &0.4094 &0.9117 &\textbf{0.9509}\\

~&~&NDCG@10 &0.8589 &0.7704 &0.7759 &\textit{0.8836 }&0.1597 &0.8796 &\textbf{0.9103}\\

~&~&MRR@10 &0.8384 &0.7370 &0.7308 &\textit{0.8675}&0.0864 &0.8688 &\textbf{0.8973}\\
\midrule

\multirow{8}{*}{\rotatebox{90}{Specified Bottom}} & \multirow{4}{*}{\rotatebox{90}{Polyvore}}
& AUC &0.6488 &\textit{0.7493} &0.7368 &0.7123 &0.7006 &0.7231 &\textbf{0.7855}\\

~&~&HR@10 &0.2092 &\textit{0.3594} &0.2613 &0.2879 &0.3065 &0.3094 &\textbf{0.4880} \\

~&~&NDCG@10 &0.1033 &\textit{0.2023} &0.1311 &0.1486 &0.1810 &0.1649 &\textbf{0.3302}\\
~&~&MRR@10 &0.0716 &\textit{0.1546} &0.0917 &0.1066 &0.1432 &0.1212 &\textbf{0.2818}\\
\cmidrule(lr){2-10}
~&\multirow{4}{*}{\rotatebox{90}{IQON}}
&AUC &0.7924 &\textit{0.8585} &0.8446 &0.8576 &0.6459 &0.8091 & \textbf{0.9016}\\
~&~&HR@10&0.6629 &0.7559 &0.7208 &\textit{0.7810} & 0.4169 &0.6705 &\textbf{0.8588} \\
~&~&NDCG@10&0.5094 &0.6232 &0.5550 &\textit{0.6743} &0.2205 &0.4931 &\textbf{0.7992} \\
~&~&MRR@10 &0.4606 &0.5818 &0.5025 &\textit{0.6402} &0.1609 &0.4378 & \textbf{0.7804}\\
\hline
\end{tabular}\\
\end{center}
\label{tab: ablation}
\end{table*}

\subsection{Ablation Study (\textbf{RQ2})}
To systematically evaluate the impact of the four key modules integrated into our proposed APCL framework, we performed ablation studies on variants of the model, specifically APCL-w/o-AP, APCL-w/o-CL, APCL-w/o-P, and APCL-w/o-C. 

These variants were created by removing the Adaptive Preference (AP), Cross-View Contrastive Learning (CL), Personal Preference (P), and Product Compatibility (C) modules from the APCL framework. More specifically, APCL-w/o-CL means our proposed method without all contrastive learning loss but only keeps the BPR loss (i.e. keeping only \cref{eq.bprloss} in \cref{eq.allloss}).
Additionally, to evaluate the individual contribution of visual and textual modalities, we conducted experiments on APCL-w/o-V and APCL-w/o-T models across both datasets. The notation -w/o-V represents a model that retains only textual information while excluding visual data, and vice versa for -w/o-T, which includes only visual information and omits textual data.

\Cref{tab: ablation} presents a comprehensive comparison of the performance between the APCL model and its variants, measured across four metrics. 
The ablation results provide strong evidence for the central design principles of APCL.

First, removing the AP module leads to consistent performance degradation across all datasets and metrics, confirming that explicit indirect relational learning contributes substantially to recommendation quality.

Second, removing the functional-view contrastive learning module also reduces performance, suggesting that alignment between direct and indirect relational views improves representation consistency.

Third, the degradation observed after removing either the personalization or compatibility module demonstrates that both relational dimensions remain essential for personalized fashion matching.

Overall, the results indicate that explicit indirect relational views and their contrastive alignment jointly contribute to the effectiveness of APCL.

\subsection{Contrastive Learning Analysis (\textbf{RQ2})}

\begin{figure}[h]
  \centering
  \setlength{\abovecaptionskip}{0.2cm}
  \includegraphics[width=\linewidth]{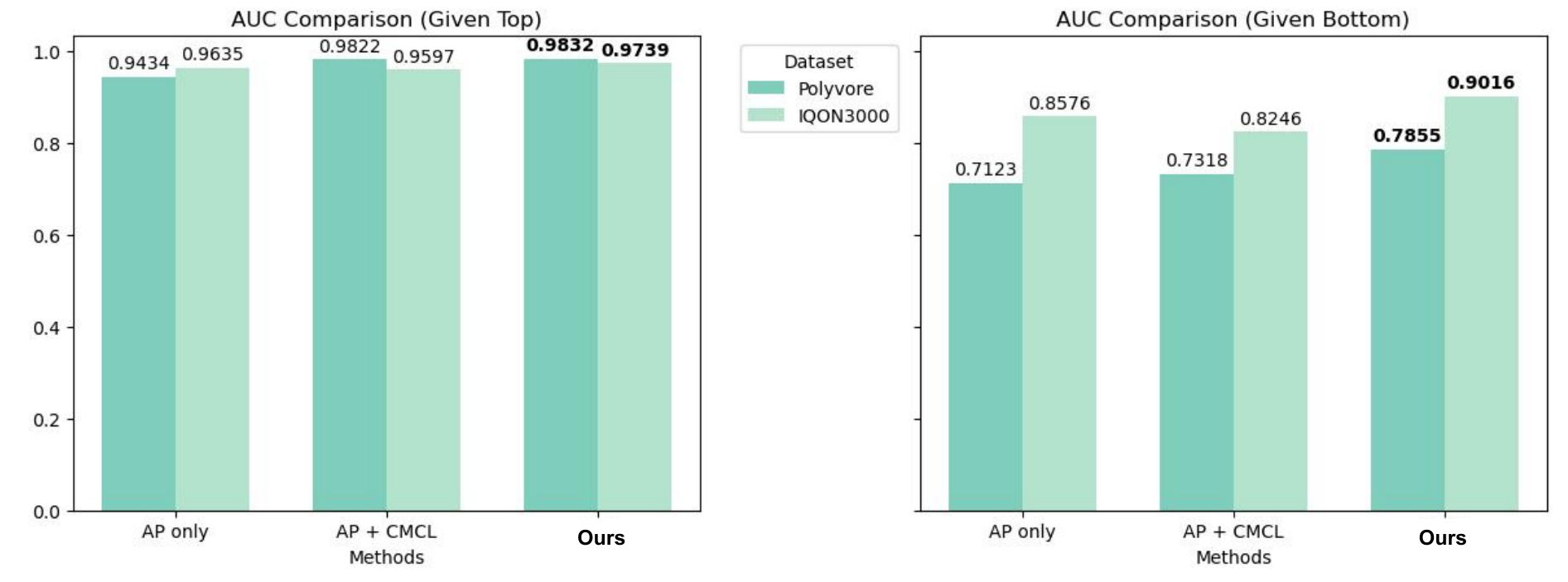}
  \caption{Comparison on Corss Modality contrastive learning(CMCL) and Corss IP and IC contrastive learning(CL) on two datasets in terms of AUC.}
  \label{fig: ablation_CL}
\end{figure}

In \Cref{fig: ablation_CL}, we evaluate the effectiveness of different contrastive learning strategies on two datasets. The methods assessed include our baseline model without contrastive learning (AP only, without CL); a traditional method utilizing Cross Visual-Textual Modality Contrastive Learning (AP + CMCL) within P and C modules, and our proposed method employing cross functional P-IP view and C-IC view Contrastive Learning (APCL).

The purpose of this experiment is not merely to evaluate contrastive learning itself, but to examine whether relational-view alignment is more beneficial than traditional modality alignment.

The results show that the proposed functional-view contrastive learning strategy consistently outperforms conventional cross-modality contrastive learning. This suggests that the effectiveness of contrastive learning depends not only on the existence of multiple views, but also on whether these views correspond to meaningful relational semantics.
By aligning direct and indirect relational representations rather than visual and textual modalities, APCL better preserves the structural information underlying recommendation decisions.

\subsection{Effect of Key Hyperparameters (\textbf{RQ3})}
\begin{figure}[t]
  \centering
  \setlength{\abovecaptionskip}{0.2cm}
  \includegraphics[width=.95\linewidth]{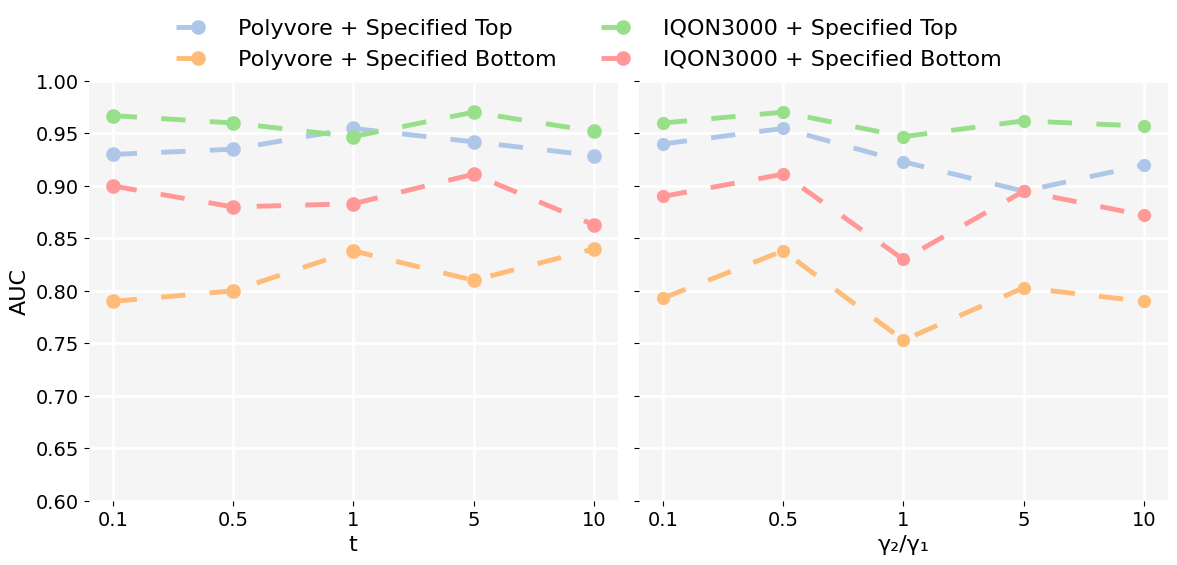}
  \caption{Impacts of different parameters.}
  \label{fig: hyper}
\end{figure}

We conduct a series of experiments to evaluate the impact of various key hyperparameters on the performance of our proposed APCL model. These hyperparameters include the temperature parameter $t$ in~\cref{eq_cl1,eq_cl2}, and the ratio $\gamma_2/\gamma_1$ in~\cref{eq.allloss}. As illustrated in~\Cref{fig: hyper}, each of these components plays a crucial role in enhancing the personalized fashion complementary recommendation performance. Notably, the model's performance exhibits only minor sensitivity to changes in these hyperparameters.
The temperature parameter $t$ is particularly influential; a higher value encourages exploration, helping the model avoid local optima, while a lower value promotes exploitation, increasing the model's confidence in differentiating between positive and negative samples. Our findings, depicted in~\Cref{fig: hyper}, indicate that the APCL model achieves optimal performance with $t$ values ranging from 1 to 5. This range fosters a more exploratory behavior, as the model becomes more uncertain and tends to assign similar probabilities to both positive and negative samples.
Furthermore, when the BPR loss and contrastive loss are given equal importance, the APCL model demonstrates its best performance. This balance appears to be critical in optimizing the model's effectiveness.

\subsection{Long Tail/Sparsity Effectiveness (\textbf{RQ4})}
\begin{figure}[ht]
  \centering
  \setlength{\abovecaptionskip}{0.2cm}
  \includegraphics[width=0.95\linewidth]{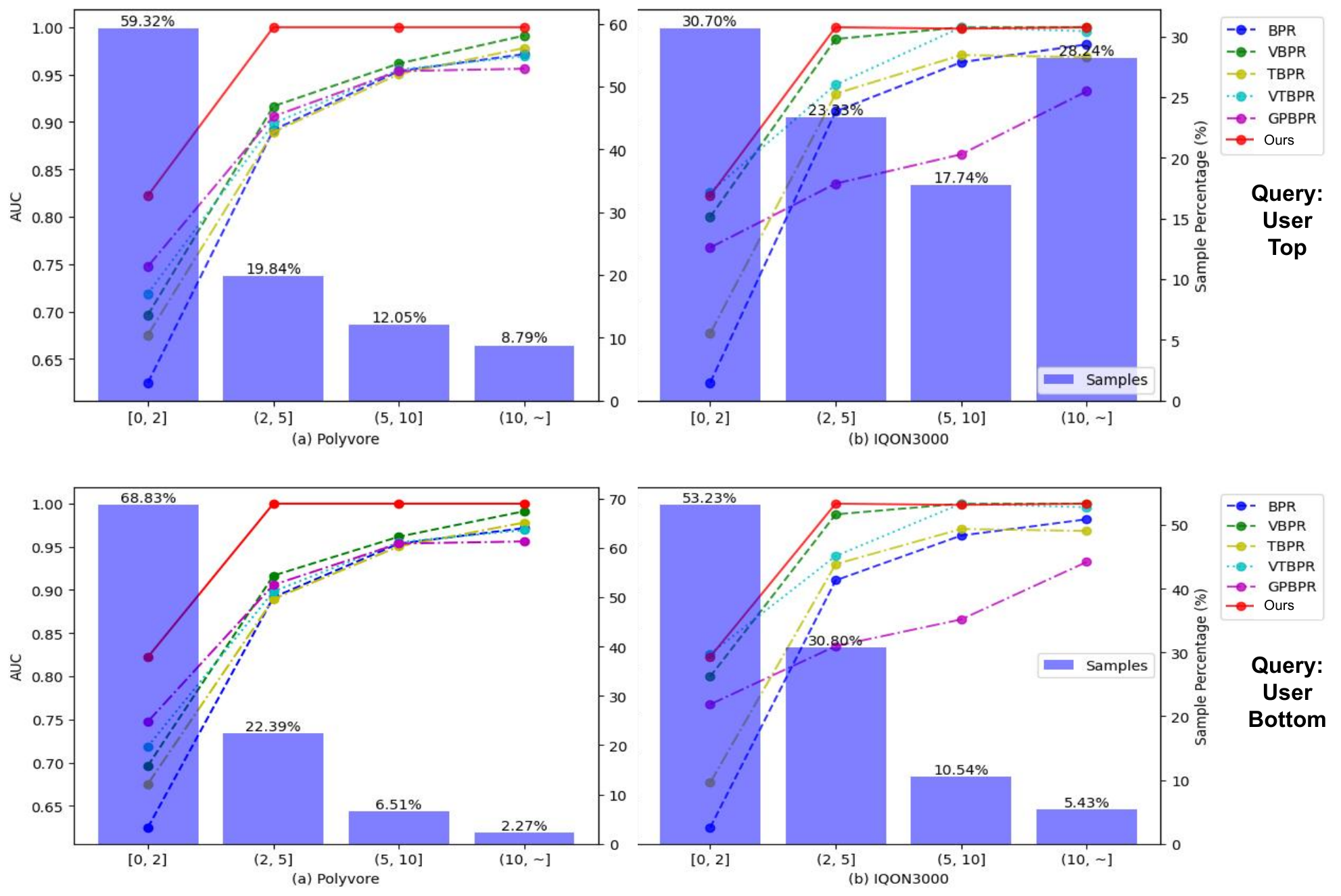}
  \caption{Comparison of Prediction Accuracy across Different Interaction Frequency Ranges for Various Methods.}
  \label{fig: long tail}
\end{figure}

To verify the effectiveness and robustness of our APCL model, we conducted a series of experiments comparing the prediction performance of various baseline models across different interaction environments. \Cref{fig: long tail} illustrates the prediction accuracy of these baselines within distinct interaction frequency ranges for the target matching item: [0, 2], (2, 5], (5, 10], and (10, ]. For instance, the range (10, ] indicates that the target matching item has more than ten interactions. Additionally, the figure includes purple bars representing the percentage of samples within each frequency range.

The sample distribution reveals that the majority of samples fall within the [0, 2] range, accounting for the majority of the total samples in all datasets. As the interaction frequency increases, the number of samples decreases, highlighting a skewed distribution where most predictions occur in the low interaction frequency range.

Despite starting with a slightly lower initial accuracy, the APCL model outperforms all other methods. It quickly achieves highest accuracy with various ranges of interaction frequency, demonstrating its robustness and efficiency. 
These findings provide empirical evidence supporting the proposed explicit indirect relational learning paradigm.

When interaction frequency is low, direct interaction signals alone are insufficient for learning reliable representations. APCL compensates for this limitation by leveraging explicit indirect relationships derived from correlated users and products. Consequently, the framework remains effective even in sparse regions of the interaction space where many existing methods struggle.

This observation is consistent with the primary motivation of the proposed framework and highlights the practical value of explicit indirect relational learning for long-tail recommendation.

\subsection{Cold-start Effectiveness (\textbf{RQ4})}
To evaluate the efficacy of our proposed indirect personal preference and product compatibility modeling for address cold-start issue, we conducted comprehensive comparative experiments against state-of-the-art baselines. The experiments were designed to isolate the impact of our indirect connection enriched contrastive learning modules by comparing model performance with and without its integration. 

\begin{figure}[t]
  \centering
  \includegraphics[width=0.95\linewidth]{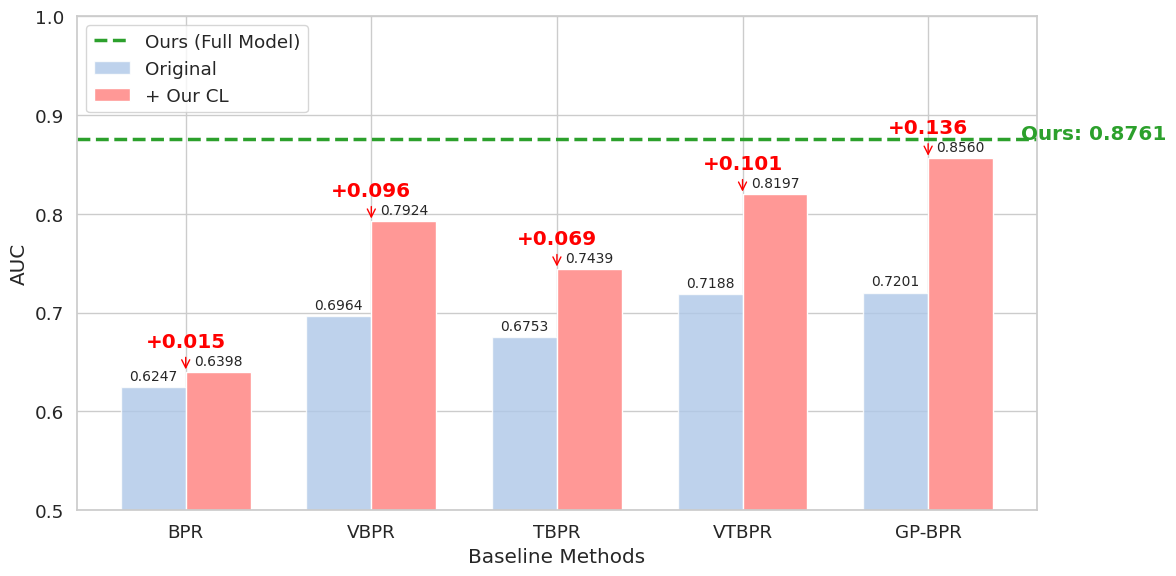}
  \caption{Comparison of prediction accuracy on cold-start with and without our CL method.}
  \label{fig: adaptive}
\end{figure}

The cold-start experiment, as illustrated in~\Cref{fig: adaptive}, directly evaluates the effectiveness of explicit indirect relational learning.

In cold-start settings, direct interaction evidence is inherently limited. Therefore, recommendation quality depends heavily on the model's ability to exploit auxiliary information. The proposed framework addresses this challenge by explicitly constructing indirect relational views from correlated users and products.

The observed improvements suggest that indirect relationships provide useful supervisory signals even when direct observations are sparse. These findings further support the hypothesis that explicit indirect relational learning can improve recommendation robustness under challenging interaction conditions.

\subsection{Qualitative Case Study (\textbf{RQ1})}
\begin{figure}[t]
  \centering
  \includegraphics[width=0.95\linewidth]{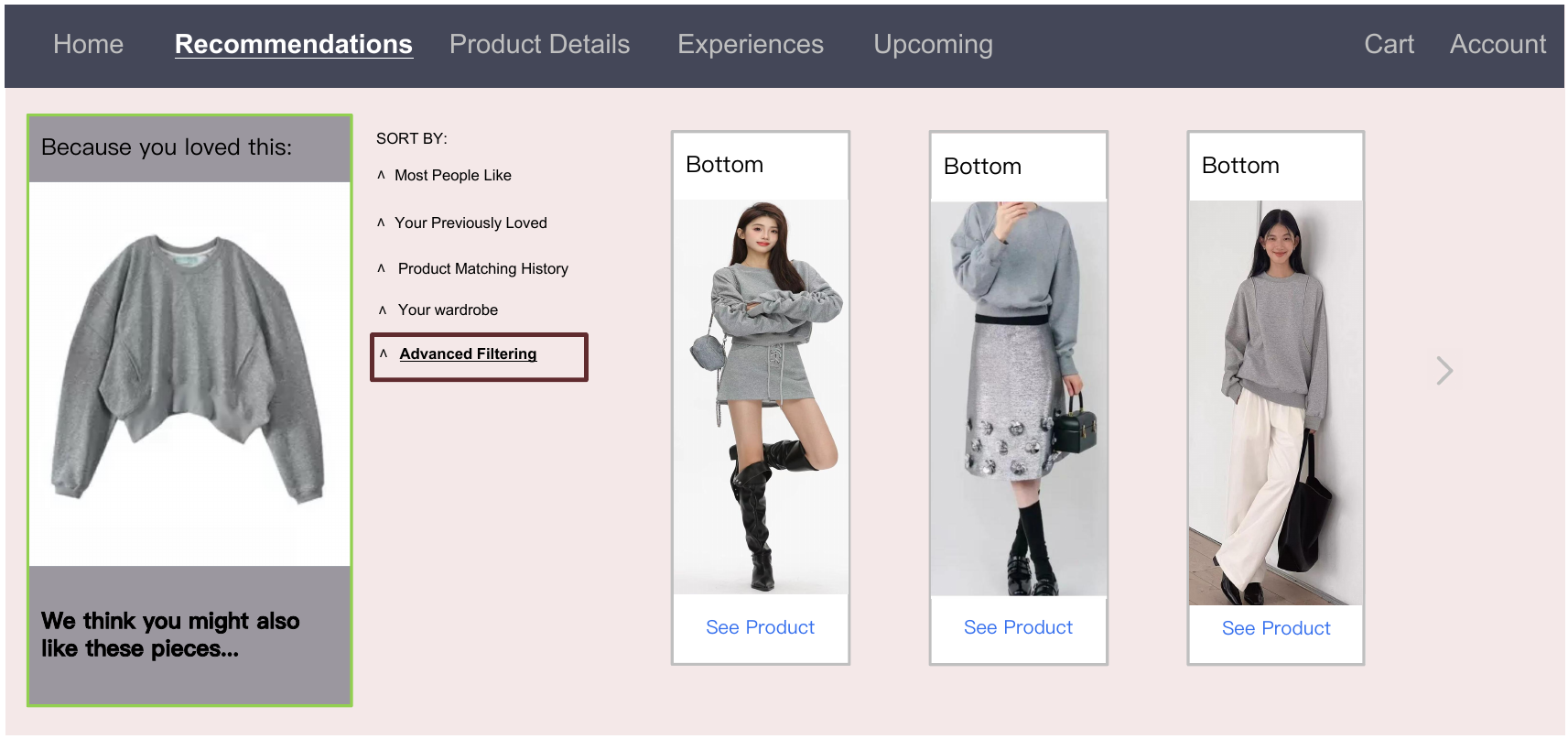}
  \caption{Potential application case.}
   \label{fig: application}
\end{figure}

\begin{figure}[h]
  \centering
  \includegraphics[width=\linewidth]{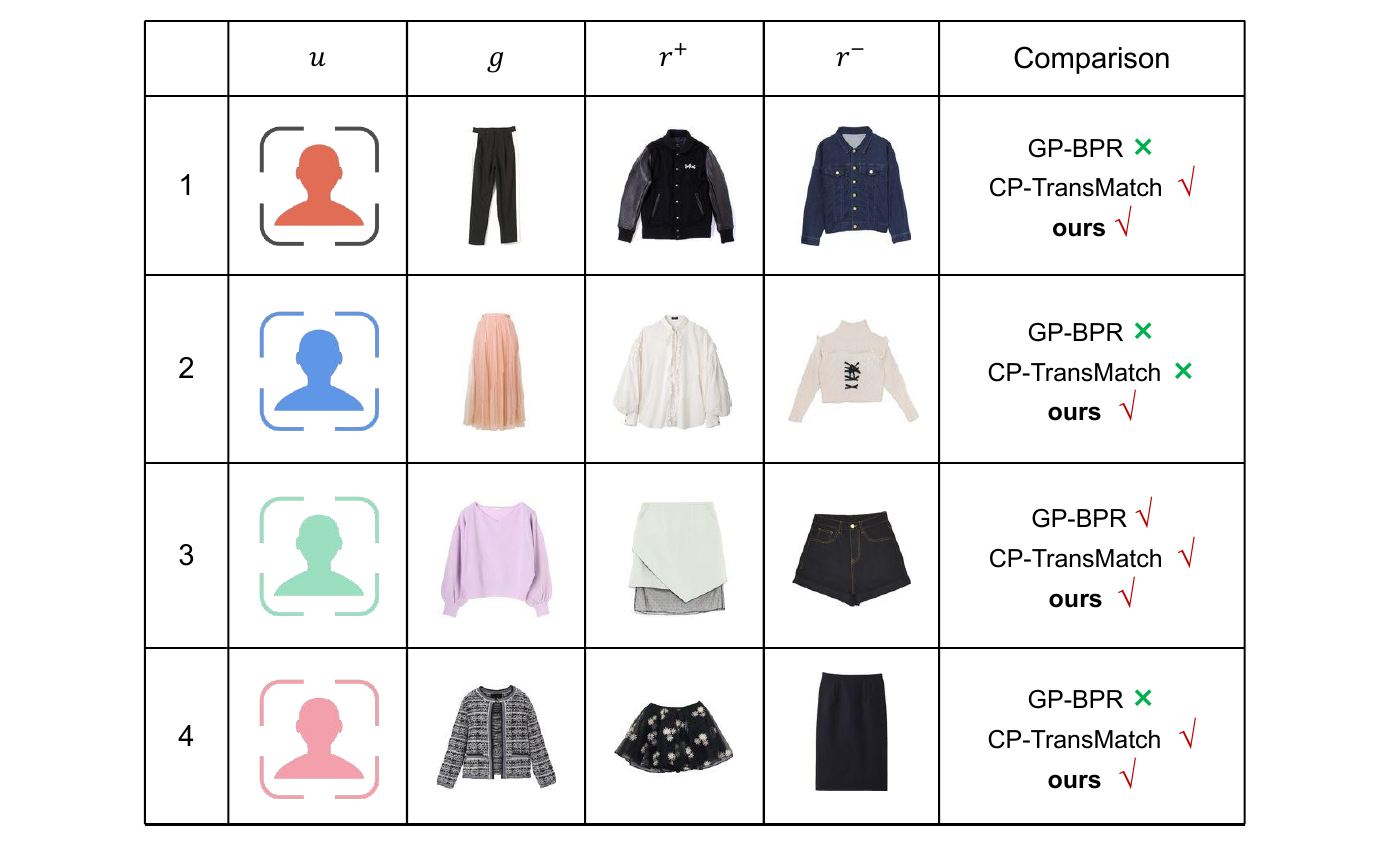}
  \caption{Personalized fashion complementary recommendation case
 results provided by different methods.}
  \label{fig: case}
\end{figure}

To verify the practical application value of our proposed method, we conducted qualitative evaluations by retrieving matching products for comparison and visualizing some test examples in~\Cref{fig: case}, and gave a potential application case in~\Cref{fig: application}. 
Specifically, in~\Cref{fig: case}, we randomly selected several user-item-match transaction triplets, denoted as $<u, g, r_+>$, from the testing dataset. In each triplet, we chose a negative matching product ($r_-$) by selecting a product that the user had not interacted with before. The objective was to correctly match the appropriate bottom clothing ($r_+$) for each user given the corresponding top garment ($g$).
As shown in in~\Cref{fig: case}, our model demonstrates strong performance across all cases. 

A closer inspection of the case studies reveals that APCL is particularly effective when several candidate products exhibit similar visual or textual characteristics.

In such situations, compatibility-based methods often struggle to distinguish among multiple plausible candidates because direct product features alone provide limited discrimination. By incorporating explicitly modeled indirect relationships, APCL is able to leverage additional contextual information originating from correlated users and products.

These examples illustrate how explicit indirect relational learning contributes beyond conventional compatibility modeling and provides more personalized recommendation outcomes.


\subsection{Discussion}
Across all experimental settings, the results consistently support the central hypothesis of this work: explicitly modeling indirect relational signals improves recommendation quality under sparse interaction conditions.

Unlike graph-based approaches that capture indirect relationships implicitly through propagation, APCL explicitly constructs indirect user--item and item--item relational views and aligns them through contrastive learning. The empirical results suggest that such explicit relational modeling provides complementary information that cannot be fully recovered from direct interactions alone.

Nevertheless, the effectiveness of the proposed approach depends on the availability of sufficiently informative correlated users or products. In extremely sparse environments where such correlations are weak or unavailable, the benefits of indirect relational learning may be reduced. Addressing this limitation represents an important direction for future research.

\section{Conclusion and Future Work}
\label{sec:concl}
This paper presented APCL, an adaptive preference modeling framework based on explicit indirect relational learning for personalized fashion matching. The central motivation of this work is that many informative recommendation signals originate from indirect relationships among users and products, particularly in sparse interaction environments. Unlike existing approaches that primarily capture such signals implicitly through graph propagation or co-occurrence statistics, APCL explicitly constructs indirect user--item and item--item relationships and represents them as dedicated relational views.

To effectively exploit these relational signals, the proposed framework decomposes recommendation into two complementary dimensions, namely personalization and compatibility, and further enriches both through indirect relational modeling. In addition, a functional-view contrastive learning mechanism is introduced to align direct and indirect representations within both relational spaces. By treating indirect relationships as explicit learning objectives rather than latent by-products of representation learning, APCL provides a more structured and controllable approach to modeling higher-order interactions.

Experimental results on the Polyvore and IQON3000 benchmark datasets demonstrate the effectiveness of the proposed framework. The consistent improvements observed across multiple evaluation metrics suggest that explicit indirect relational learning can successfully complement direct interaction modeling. Further analyses under sparse-data, long-tail, and cold-start settings show that indirect relational signals provide valuable auxiliary information when direct observations are insufficient, highlighting the practical advantages of the proposed approach.

Despite these promising results, several limitations remain. First, the current study focuses on pairwise fashion matching involving top--bottom combinations and does not consider more complex outfit compositions. Second, the effectiveness of explicit indirect relational learning depends on the availability of meaningful correlated users or products; therefore, its benefits may diminish in extremely sparse environments where such relational evidence is limited. Finally, the introduction of indirect relational construction and aggregation mechanisms incurs additional computational overhead compared with simpler recommendation models.

Future work will investigate extending the proposed framework to multi-category outfit recommendation involving shoes, accessories, and other fashion products. We also plan to explore richer contextual signals, such as temporal dynamics, session behaviors, and user intent, as additional sources of indirect relational information. Another promising direction is the integration of explicit indirect relational learning with graph-based recommendation architectures, combining the advantages of explicit relational reasoning and graph propagation. Furthermore, generative recommendation techniques, such as diffusion-based outfit synthesis, offer opportunities for leveraging indirect relational signals in fashion content generation and recommendation simultaneously.

Overall, this work introduces explicit indirect relational learning as a new perspective for personalized fashion matching. The results demonstrate that explicitly constructing, modeling, and aligning indirect relational views can improve recommendation robustness and effectiveness, providing a structured alternative to purely implicit relational modeling approaches and offering potential value for a broader class of recommendation problems beyond the fashion domain.

\section*{Acknowledgments}
The work described in this paper was supported, in part, by the Innovation and Technology Fund (Project: ITP/004/24TP) and by The Hong Kong University of Science and Technology (Grant: R9973). 

\section*{CRediT authorship contribution statement}
\textbf{Shuiying Liao:} Conceptualization, Methodology, Formal analysis, Writing - original draft. 
\textbf{Li Li:} Supervision, Visualization.
\textbf{P.Y. Mok:} Methdology, Formal analysis, Supervision, Funding acquisition, Writing - review \& editing.



\section*{Data availability statement}
The data supporting the findings of this study are openly available from the original references listed in the article.  The code is available at \url{https://anonymous.4open.science/r/APCL-05BD}.





 \bibliographystyle{elsarticle-num} 
 \bibliography{APCL.bib}






\end{document}